\documentclass[10pt,aps,prc,twocolumn,tightenlines,showpacs,floatfix]{revtex4-1}

\usepackage{dcolumn}
\usepackage{xcolor}
\usepackage{bm}
\usepackage{graphicx}
\usepackage{hyperref}

\definecolor{linkcolor}{rgb}{0,0,0.40} 
\hypersetup{
    pdfsubject=Paper,
    pdfkeywords={nuclear physics} {Bayesian} {pionless EFT} {discrepancy model},
    unicode = true,
    breaklinks = true,
    colorlinks = true,
    linkcolor = linkcolor,
    citecolor = linkcolor,
    menucolor = linkcolor,
    urlcolor = linkcolor
}

\usepackage{amsmath,amsfonts,amssymb}
\usepackage{slashed}
\usepackage{verbatim}

\newcommand{\taudot}{{\bm \tau}_1 \cdot {\bm \tau}_2}
\newcommand{\sigmadot}{{\bm \sigma}_1 \cdot {\bm \sigma}_2}
\newcommand{\cbar}{\bar{c}}
\newcommand{\pionlessEFT}{$\slashed{\pi}$EFT}
\begin{document}

{
\title{Bayesian analysis of nucleon-nucleon scattering data in pionless effective field theory}
\author{J. M. Bub$^{1}$}
\author{M.\ Piarulli$^{1,2}$}
\author{R. J. Furnstahl$^{3}$}
\author{S.\ Pastore$^{1,2}$}
\author{D. R. Phillips$^{4}$}
\affiliation{
$^1$\mbox{Department of Physics, Washington University in St. Louis, St. Louis, MO 63130, USA}\\
$^2$\mbox{McDonnell Center for the Space Sciences at Washington University in St. Louis, MO 63130, USA}\\
$^3$\mbox{Department of Physics, The Ohio State University, Columbus, OH 43210, USA}
$^4$\mbox{Institute of Nuclear and Particle Physics and Department of Physics and Astronomy, Ohio University, Athens, OH 45701, USA}\\
}

\begin{abstract}
We perform Bayesian model calibration of two-nucleon ($NN$) low-energy constants (LECs) appearing in an $NN$ interaction based on pionless effective field theory ($\slashed{\pi}$EFT). The calibration is carried out for potentials constructed using naive dimensional analysis in $NN$ relative momenta ($p$) up to next-to-leading order [NLO, $O(p^2)$] and next-to-next-to-next-to-leading order [N3LO, $O(p^4)$]. We consider two classes of $\slashed{\pi}$EFT potential: one that acts in all partial waves and another that is dominated by $s$-wave physics. The two classes produce broadly similar results for calibrations to $NN$ data up to $E_{\rm lab}=5$ MeV. Our analysis accounts for the correlated uncertainties that arise from the truncation of the $\slashed{\pi}$EFT. We simultaneously estimate both the EFT LECs and the parameters that quantify the truncation error. This permits the first quantitative estimates of the $\slashed{\pi}$EFT breakdown scale, $\Lambda_b$: the 95\% intervals are $\Lambda_b \in [50.11,63.03]$ MeV at NLO and $\Lambda_b \in [72.27, 88.54]$ MeV at N3LO. Invoking naive dimensional analysis for the $NN$ potential, therefore, does not lead to consistent results across orders in $\slashed{\pi}$EFT. This exemplifies the possible use of Bayesian tools to identify inconsistencies in a proposed EFT power counting. 
\end{abstract}

\maketitle

}

\section {Introduction}
\label{sec:intro}

One of the key goals of nuclear physics is the quest to understand nuclear phenomena based on the microscopic interactions of the constituent nucleons.
Over the past three decades, significant progress in the field has been driven by the development of powerful {\it ab initio} many-body methods~\cite{Hergert:2020bxy, Ekstrom:2022yea, Machleidt:2023jws,Navratil:2022lvq} for approximately solving the nuclear Schrödinger equation and the formulation of nuclear forces using Effective Field Theory (EFT).
A more recent development has been the application of Bayesian methods~\cite{Furnstahl:2014xsa,Phillips:2020dmw}, which help combine prior knowledge with experimental data, enabling more accurate estimates of uncertainties and parameter distributions. Bayesian methods, when applied to EFTs, provide a robust framework for developing reliable and predictive nuclear models~\cite{Wesolowski:2021cni}. 

The primary objective of this paper is to initiate a project to integrate and advance the Bayesian methodologies developed by the BUQEYE (Bayesian Uncertainty Quantification: Errors for Your EFT)~\cite{BUQEYE,Schindler:2008fh,Wesolowski:2018lzj, Melendez:2019izc, Wesolowski:2021cni} and BAND (Baysian Analysis of Nuclear Dynamics)~\cite{bandframework} collaborations, adapting them specifically to local EFT nucleon-nucleon ($NN$) interactions in coordinate space. 
These interactions serve as essential inputs for Quantum Monte Carlo (QMC) methods~\cite{Gandolfi:2020pbj,Carlson:2014vla}, which are instrumental in accurately determining the properties of light nuclei—ranging from their spectra and moments to transitions~\cite{Piarulli:2017dwd,Baroni:2018fdn,King:2020pza,King:2020wmp,Cirigliano:2019vdj,Cirigliano:2018hja,King:2022zkz,King:2021jdb,Piarulli:2022ulk,Chambers-Wall:2024fha,Chambers-Wall:2024uhq,Andreoli:2024ovl}—as well as determining the equation of state of neutron and nuclear matter~\cite{Piarulli:2019pfq,Lovato:2022apd,Fore:2022ljl}. Therefore, this study represents a foundational step toward rigorous theoretical uncertainty quantification in QMC calculations.

The EFT paradigm, which emerged in the early 1990s~\cite{Weinberg:1990rz,Weinberg:1991um,Weinberg:1992yk,Ordonez:1992xp,Rho:1990cf}, has provided the means for the systematic expansion of inter-nucleon interactions with a power counting that dictates what mechanisms appear at each order. An important component of this work involves applying Bayesian tools to analyze and interpret $NN$ scattering data with respect to the EFT model under consideration, conditioned by a specific power counting scheme. We are interested in quantitatively evaluating the model's performance by incorporating both experimental and theoretical uncertainties in the model calibration, shedding light on its predictive strengths and potential limitations, including aspects of power counting and the renormalization scheme.

A crucial point of any EFT is the choice of degrees of freedom (dofs), which for nuclear physics can include
nucleon-only interactions, called pionless interactions; pion-full interactions, known as chiral interactions; and $\Delta$-full interactions, where $\Delta$-isobar states are included as intermediate states. 
Two-nucleon chiral interactions have been constructed to high order in the low-energy expansion~\cite{Ordonez:1993tn,vanKolck:1994yi,Ordonez:1995rz,Entem:2003ft,Epelbaum:2004fk,Ekstrom:2013kea,Entem:2014msa,Epelbaum:2014sza,Gezerlis:2014zia,Piarulli:2014bda,Entem:2015xwa,Ekstrom:2015rta,Piarulli:2016vel,Entem:2017gor,Reinert:2017usi,Ekstrom:2017koy,Jiang:2020the,Saha:2022oep,Somasundaram:2023sup} and provide a precise description of $NN$ scattering data up to the pion production threshold (see Ref.~\cite{Machleidt:2024bwl} for a comprehensive review of the recent advances in chiral EFT ($\chi$EFT) based nuclear forces and their applications).  
However, we begin our exploration of Bayesian tools here with pionless EFT ($\slashed{\pi}$EFT), which has been successfully applied to various problems involving light and medium-mass nuclei; representative applications can be found in Ref.~\cite{Hammer:2019poc}.

Our choice of starting with $\slashed{\pi}$EFT is a pragmatic one. As the interaction is low-energy and $s$-wave dominant, $\slashed{\pi}$EFT yields a simpler interaction than its chiral counterpart. As no light mesons mediate the interaction in $\slashed{\pi}$EFT, its interaction is characterized solely by short-range interactions. However, the relative simplicity of $\slashed{\pi}$EFT does not diminish the value of exploring it within the BUQEYE framework. Although numerous studies have utilized these frameworks on various chiral models~\cite{Wesolowski:2018lzj,Wesolowski:2021cni,Svensson:2023twt,Thim:2023fnl,Somasundaram:2023sup}, to the best of our knowledge, no such analysis has been performed to date on $\slashed{\pi}$EFT.
Since the BUQEYE framework aims at providing a general approach to quantifying EFT uncertainties, applying it to $\slashed{\pi}$EFT allows for further validation of its methods. Furthermore, $\slashed{\pi}$EFT has a generally universally accepted power counting~\cite{Kaplan:1998tg,Kaplan:1998we,vanKolck:1998bw,Birse:1998dk}. This fact makes $\slashed{\pi}$EFT an excellent system to reconcile the conflicts of power counting of EFTs when treated in a rigorous and consistent manner. In particular, the introduction of Bayesian methods from the BUQEYE collaboration may be well suited for this task. 

The BUQEYE framework allows us to investigate other properties inherent to EFTs, namely the EFT breakdown scale, $\Lambda_b$. The choice of EFT dofs dictates $\Lambda_b$, which delineates the boundary between the domain of applicability and high-energy processes that the EFT does not resolve. The breakdown scale arises due to the exclusion of massive degrees of freedom in the EFT, which are integrated out of the interaction. All of the effects from the integrated-out degrees of freedom are encapsulated in low-energy constants (LECs), which characterize the short-range interaction. 
In $\chi$EFT, all but the lightest meson, the pion, are integrated out, which sets the breakdown scale of order the mass of the lightest non-Goldstone mesons.
In $\slashed{\pi}$EFT, pions are integrated out of the interaction, leading to a breakdown scale set below the pion mass, $m_{\pi}$~\cite{Bedaque:1997qi, Bedaque:1998mb, Kaplan:1998tg, Kaplan:1998we, vanKolck:1998bw, Birse:1998dk,Chen:1999tn}.
By analyzing the order-by-order convergence of $\slashed{\pi}$EFT, we can verify the expectation for $\Lambda_b$ to be $m_{\pi}$.

Additionally, when working with EFTs, the expansion is truncated at some order which induces uncertainties from neglected physics. These must be accounted for to enable controlled predictions of nuclear observables.
We use Bayesian methods to help manage and quantify these uncertainties, facilitating the quantification of precision nuclear physics and enabling the most meaningful comparisons between theoretical predictions and experimental or observational data.

Specifically, we apply the Bayesian methodology to two classes of $\slashed{\pi}$EFT at next-to-leading order (NLO) and next-to-next-to-next-to-leading order (N3LO) following naive dimensional analysis, commonly referred to as Weinberg power counting, which we discuss in Sec.~\ref{subsec:contact}. 
The first class contains the full local operator structure, and we label this as $\slashed{\pi}_{\text{WB}}$(NLO/N3LO)$_{\text{Full}}$ while the second one contains a subset of these operators that capture predominately $s$-wave physics, which we label as $\slashed{\pi}_{\text{WB}}$(NLO/N3LO)$_{\text{Red.}}$. These interactions have been regularized with coordinate-space Gaussian regulators, with cutoffs of $R_s=[1.5,2.0,2.5]$ fm. Further, in our nomenclature, we will make note of the cutoff at the end of our naming scheme, for example, $\slashed{\pi}_{\text{WB}}$NLO$_{\text{Full}}$-2.0. 

Although this work focuses on the BUQEYE framework in the Weinberg scheme, where the size of the LECs in the interaction aligns with naive dimensional analysis~\cite{Manohar:1983md,Georgi:1992dw}, the analysis is not limited to only this scheme. 
In future work, we plan to apply this framework to other power countings, including the accepted ``standard" power counting of $\slashed{\pi}$EFT~\cite{Kaplan:1998tg,Kaplan:1998we} that enhances the importance of certain LECs. Additionally, we aim to extend our analysis to local chiral interactions, developed in Refs.~\cite{Piarulli:2014bda,Piarulli:2016vel,Baroni:2018fdn}.

The paper is organized as follows. In Sec.~\ref{sec:int}, we detail the $NN$ interaction under consideration, with an emphasis on projecting the LECs into specific spin and isospin channels, $S=\{0,1\}$ and $T=\{0,1\}$ respectively. In Sec.~\ref{sec:methods} we outline the Bayesian framework developed by the BUQEYE collaboration, applying it to calibrate the LECs in our $\slashed{\pi}$EFT models. 
Sec.~\ref{sec:results} presents the results of our analysis discussing the magnitude of parametric errors in the model calibration, the assessment of truncation errors, the convergence of the LECs with respect to these errors, and the estimation of the breakdown scale, $\Lambda_b$. It also explores the application of posterior distributions to estimate properties of the deuteron and the effective range expansion parameters for proton-proton ($pp$) and neutron-proton ($np$) scattering. The paper concludes with Sec.~\ref{sec:summary}, where we summarize our main findings and discuss directions for future research.

\section{Interaction}
\label{sec:int}

\subsection{Contact interactions}
\label{subsec:contact}
The nucleon-nucleon ($NN$) interaction under consideration adheres to a momentum hierarchy defined by
\begin{equation}\label{eq:eft_expansion_parameter}
    Q \equiv \frac{\mathrm{max}[p_{\mathrm{soft}},P]}{\Lambda_b} < 1,
\end{equation}
where $P$ is the characteristic momentum of the process under consideration and $p_{\mathrm{soft}}$ is the essential low momentum scale of the physics captured by the EFT. The momenta involved are ${\bf q} = {\bf p} - {\bf p'}$ and ${\bf k} = \frac{{\bf p} + {\bf p'}}{2}$, representing the momentum transfer and the center of mass momentum, respectively. Here, ${\bf p}$ and ${\bf p'}$ are the initial and final relative momenta of the two nucleons, respectively. This analysis follows the contact interaction order established by naive dimensional analysis, i.e., Weinberg power-counting, in scenarios involving explicit pion fields, as detailed in Refs.~\cite{Weinberg:1990rz,Weinberg:1991um,Weinberg:1992yk}. We further choose to maintain the same labeling of orders that would be present in a $\pi$-full EFT, where the orders that we work at are: leading order (LO or $Q^0$), next-to-leading order (NLO or $Q^2$), and N3LO ($Q^4$). These interactions consist of charge-independent (CI) terms at LO, NLO, and N3LO,
and charge-dependent (CD) ones at NLO and N3LO.

First, at LO, the CI interaction has zero derivative operators of the form
\begin{equation}
    v^{\rm CI}_{\rm LO} = C_S + C_T \sigmadot,
\end{equation}
where $\sigma_i$ denotes the Pauli spin operator.

At NLO, the interaction has the added two-derivative CI contribution of the form
\begin{align}
    v^{\rm CI}_{\rm NLO}({\bm q},{\bm k}) = &C_1 q^2 + C_2 q^2 \taudot + C_3 q^2 \sigmadot \nonumber \\ 
    &+ C_4 q^2 \sigmadot \taudot + C_5 S_{12}({\bf q}) \nonumber \\ &
    + C_6 S_{12}({\bf q}) \taudot + i C_7 {\bf S \cdot (k \times q)}
\end{align}
where $\tau_i$ denotes the Pauli isospin operator and $S_{12}({\bm q}) = 3 \left({\bm \sigma}_1 \cdot {\bm q} \right) \, \, \left({\bm \sigma}_2 \cdot {\bf q} \right) - q^2 \left(\sigmadot \right)$ is the tensor operator. Also, at NLO, we introduce a CD interaction
\begin{equation}
    v^{\rm CD}_{\rm NLO} = C_0^{\rm IT} T_{12} + C_0^{\rm IV}(\tau_{1z} + \tau_{2z}),
\end{equation}
where $T_{12} = 3 \tau_{1z}\tau_{2z} - \taudot$ and $\tau_{1z} + \tau_{2z}$ are the isotensor (IT) and isovector (IV) operators, respectively. Any calculation that we perform at NLO employs a total interaction of 
\begin{equation}v({\bm q},{\bm k}) = v^{\rm CI}_{\rm LO} + v^{\rm CI}_{\rm NLO}({\bm q},{\bm k}) + v^{\rm CD}_{\rm NLO}.
\end{equation}

Finally, at N$^3$LO the CI interaction takes one final correction with four derivative operators
\begin{align}
\label{eq:vCI_N3LO}
    v^{\rm CI}_{\rm N3LO}({\bm q},{\bm k}) &= D_1 q^4 + D_2 q^4 \taudot + D_3 q^4 \sigmadot \nonumber \\
    & \qquad\null + D_4 q^4 \sigmadot \taudot + D_5 q^2 S_{12}({\bm q}) \nonumber \\
    &\qquad\null + D_6 q^2 S_{12}({\bm q}) \taudot + iD_7 q^2 {\bf S \cdot (k \times q)} \nonumber \\
    & \qquad\null + iD_8 q^2 {\bf S \cdot (k \times q)}\taudot \nonumber \\
    & \qquad\null + D_9 [{\bf S \cdot (k \times q)}]^2 + D_{10} {\bf (k \times q)}^2 \nonumber \\ & \qquad\null + D_{11}{\bf (k \times q)}^2 \sigmadot.
\end{align}
At N3LO we also have additional CD corrections:
\begin{align}
    v_{\rm N3LO}^{\rm CD} ({\bf q,k}) = &[C_1^{\rm IT}q^2 + C_2^{\rm IT} q^2 \sigmadot + C_3^{\rm IT} S_{12}(\bm q) \nonumber \\
    &+ i C_4^{\rm IT} {\bf S \cdot (k \times q)} ]T_{12}.
\end{align}
Thus, our complete interaction for calculations performed at N3LO is of the form \begin{align}
v({\bm q},{\bm k}) &= v^{\rm CI}_{\rm LO} + v^{\rm CI}_{\rm NLO}({\bm q},{\bm k}) + v^{\rm CD}_{\rm NLO} \notag \\ 
  & \qquad\null + v^{\rm CI}_{\rm N3LO}({\bf q,k}) + v_{\rm N3LO}^{\rm CD} ({\bf q,k}).
\end{align}

We observe that at N3LO, four additional CI contact interaction structures that are not included in Eq.~(\ref{eq:vCI_N3LO}) are also permitted. Following the guidance of Ref.~\cite{Piarulli:2016vel}, we exclude these terms because they generate operator structures in configuration space that depend quadratically on the relative momentum operator, complicating their implementation in quantum Monte Carlo calculations~\cite{Piarulli:2017dwd}. Indeed, including these terms has not demonstrated any improvement in fitting $NN$ scattering data~\cite{Piarulli:2016vel}. Additionally, three combinations of these terms are null on the energy shell~\cite{Wesolowski:2018lzj, Reinert:2017usi}, and their effects can be incorporated through a redefinition of the $3N$ interaction ~\cite{Girlanda:2020pqn}. We also disregard five IV terms in the CD sector—one at NLO and four at N3LO~\cite{Piarulli:2016vel}. Notably, only one observable, the difference between the $pp$ and $nn$ scattering lengths, is sensitive to these terms.

We refer the reader to Refs.~\cite{Piarulli:2014bda, Piarulli:2016vel, Schiavilla:2021dun} for details on the corresponding coordinate-space representation of the interaction. Along the lines of Refs.~\cite{Piarulli:2014bda, Piarulli:2016vel, Schiavilla:2021dun}, we also include the complete electromagnetic interaction, $v^{\rm EM}$, at each order of the EFT expansion. This interaction accounts for terms up to quadratic in the fine structure
constant (first and second order Coulomb, Darwin-Foldy, vacuum polarization, and magnetic moment terms), see
Ref.~\cite{Wiringa:1994wb} for explicit expressions.

\subsection{Projected Basis}
\label{subsec:Projected Basis}
In this paper, we present the LECs projected into specific spin and isospin channels, $S=\{0,1\}$ and $T=\{0,1\}$ respectively. Thus, we use projection operators in the given channel, written as 
\begin{equation}\label{eq:projection_op}
        P_0^{\sigma} = \frac{1 - {\bm \sigma}_1 \cdot {\bm \sigma}_2 }{4} , \quad P_1^{\sigma} = \frac{3 + {\bm \sigma}_1 \cdot {\bm \sigma}_2 }{4},
\end{equation} 
and likewise for $P_0^{\tau}$ and $P_1^{\tau}$, which project on pairs with $S$ ($T$) equal 1 and 0 respectively. 

Thus, for example, at LO, our interaction in this projected basis is written as 
\begin{equation}
    v^{\rm CI}_{\mathrm{LO}} = (c0c0) P_0^{\sigma} + (c0c1) P_1^{\sigma}.
\end{equation}
These projection matrices generate the transformation from the basis in Sec.~\ref{subsec:contact} to the projected basis. The projected coefficients are denoted as $cxxxx$, where the second character indicates the order of the EFT expansion, with possible values including $0$, $2$, or $4$; the third character specifies the type of operators, such as central ($c$), central isovector ($cv$), central isotensor ($ct$), tensor ($t$), tensor isotensor ($tt$), spin-orbit ($b$), spin-orbit isotensor ($bt$), spin-orbit squared ($bb$), and orbital momentum squared ($q$). The fourth and fifth characters represent the spin and isospin projection channels, respectively. We note, however, that this naming convention regarding the fourth and fifth characters is not rigorously adhered to. Specifically, if the LEC corresponds to an operator without spin and/or isospin dependence, the fourth and/or fifth characters are omitted. Additionally, in cases where specific spin and/or isospin channels can be inferred by the operator (like $S=1$ for $b$ or $T=1$ for $ct$) these characters are likewise dropped from our notation. Tab.~\ref{tab:LECs} provides a reference for the order-by-order LECs, organized by the type of operator. It also includes the $(S,T)$ channels and corresponding partial waves where they contribute.

The transformation into the projected basis is given in the following set of equations. First, for central potentials, we have at LO
\begin{equation}
    \begin{pmatrix} 
        c0c0 \\ c0c1 
    \end{pmatrix} = 
    \begin{pmatrix}
        1 & -3 \\ 1 & 1
    \end{pmatrix}
    \begin{pmatrix}
        C_S \\ C_T
    \end{pmatrix},
\end{equation}
where the LECs have units of fm${}^2$. Note that we require the LO interactions to act only in even partial waves, by manually suppressing the channels $(S,T)=(0,0)$ and $(1,1)$.

At NLO, the central contributions are
\begin{equation}
    \begin{pmatrix} 
        c2c00 \\ c2c10 \\ c2c01 \\ c2c11
    \end{pmatrix} = 
    \begin{pmatrix}
        1 & -3 & -3 &  9 \\
        1 & -3 &  1 & -3 \\
        1 &  1 & -3 & -3 \\
        1 &  1 &  1 &  1
    \end{pmatrix}
    \begin{pmatrix}
        C_1 \\ C_2 \\ C_3 \\ C_4
    \end{pmatrix}
\end{equation}
with units of fm${}^4$, and similarly for N3LO,
\begin{equation}
    \begin{pmatrix} 
        c4c00 \\ c4c10 \\ c4c01 \\ c4c11
    \end{pmatrix} = 
    \begin{pmatrix}
        1 & -3 & -3 &  9 \\
        1 & -3 &  1 & -3 \\
        1 &  1 & -3 & -3 \\
        1 &  1 &  1 &  1
    \end{pmatrix}
    \begin{pmatrix}
        D_1 \\ D_2 \\ D_3 \\ D_4
    \end{pmatrix}
\end{equation}
with units of fm${}^6$. We further have at NLO the central isovector
\begin{equation}
    c0cv = C_0^{\mathrm{IV}}
\end{equation}
and the central isotensor
\begin{equation}
    c0ct = C_0^{\mathrm{IT}}
\end{equation}
contributions, each with units of fm${}^4$. These are accompanied by the N3LO central isotensor terms
\begin{equation}
    \begin{pmatrix}
        c2ct0 \\ c2ct1
    \end{pmatrix} =
    \begin{pmatrix}
        1 & -3 \\
        1 & 1
    \end{pmatrix}
    \begin{pmatrix}
        C_1^{\mathrm{IT}} \\ C_2^{\mathrm{IT}}
    \end{pmatrix},
\end{equation}
which have units of fm${}^6$.

Following the central terms, the tensor terms are as follows, which at NLO are
\begin{equation}
    \begin{pmatrix} 
        c2t0 \\ c2t1
    \end{pmatrix} = 
    \begin{pmatrix}
        1 & -3 \\ 1 & 1
    \end{pmatrix}
    \begin{pmatrix}
        C_5 \\ C_6
    \end{pmatrix},
\end{equation}
and similarly for N3LO,
\begin{equation}
    \begin{pmatrix} 
        c4t0 \\ c4t1
    \end{pmatrix} = 
    \begin{pmatrix}
        1 & -3 \\ 1 & 1
    \end{pmatrix}
    \begin{pmatrix}
        D_5 \\ D_6
    \end{pmatrix},
\end{equation}
which have units of fm${}^4$ and fm${}^6$ respectively. We further have the tensor isotensor term at N3LO,
\begin{equation}
    c2tt = C_3^{\mathrm{IT}}
\end{equation}
which also have units of fm${}^6$. 

The final groups of LECs are those that are dependent on angular momentum. First, we have the NLO spin-orbit contribution,
\begin{equation}
    c2b = C_7
\end{equation}
with units of fm${}^4$, followed by the N3LO contributions,
\begin{equation}
    \begin{pmatrix}
        c4b0 \\ c4b1
    \end{pmatrix} = 
    \begin{pmatrix}
        1 & -3 \\ 1 & 1
    \end{pmatrix}
    \begin{pmatrix}
        D_7 \\ D_8
    \end{pmatrix}
\end{equation}
with units of fm${}^6$. Accompanying the N3LO CI spin-orbit interaction is the spin-orbit isotensor contribution,
\begin{equation}
    c2bt = C_4^{\mathrm{IT}}.
\end{equation}
Also at N3LO, we have a spin-orbit squared contribution,
\begin{equation}
    c4bb = D_9,
\end{equation}
and, finally, the orbital angular momentum squared contributions,
\begin{equation}
    \begin{pmatrix}
        c4q0 \\ c4q1
    \end{pmatrix} = 
    \begin{pmatrix}
        1 & -3 \\ 1 & 1
    \end{pmatrix}
    \begin{pmatrix}
        D_{10} \\ D_{11}
    \end{pmatrix},
\end{equation}
where these last three N3LO operator groups all have units of fm${}^6$.

\begin{table}
   \centering
   \begin{ruledtabular}
   \begin{tabular}{llll}
       LEC & Order & $(S,T)$ & Partial Waves \\ \hline
       \multicolumn{4}{c}{Central} \\ \hline
       c0c0  & LO  & (0,1) & $^1S_0,{}^1D_2,\ldots$ \\ 
       c0c1  & LO  & (1,0) & ${}^3S_1-{}^3D_1,{}^3D_2,{}^3D_3-{}^3G_3\ldots$ \\
       c2c00 & NLO & (0,0)   & $^1P_1,{}^1F_3\ldots$ \\ 
       c2c10 & NLO & (1,0)   & ${}^3S_1-{}^3D_1,{}^3D_2,{}^3D_3-{}^3G_3\ldots$ \\
       c2c01 & NLO & (0,1)   & $^1S_0,{}^1D_2\ldots$ \\
       c2c11 & NLO & (1,1)   & ${}^3P_0,{}^3P_1,{}^3P_2-{}^3F_2\ldots$ \\
       c4c00 & N3LO &  (0,0)  & $^1P_1,{}^1F_3\ldots$ \\ 
       c4c10 & N3LO & (1,0)   & ${}^3S_1-{}^3D_1,{}^3D_2,{}^3D_3-{}^3G_3\ldots$ \\
       c4c01 & N3LO & (0,1)   & $^1S_0,{}^1D_2\ldots$ \\
       c4c11 & N3LO & (1,1)   & ${}^3P_0,{}^3P_1,{}^3P_2-{}^3F_2\ldots$ \\ \hline
       \multicolumn{4}{c}{Central Isovector} \\ \hline
       c0cv  & NLO & (0/1,1) & ${}^1S_0,{}^3P_0,{}^3P_1,{}^3P_2-{}^3F_2,{}^1D_2\ldots$ \\ \hline
       \multicolumn{4}{c}{Central Isotensor} \\ \hline
       c0ct  & NLO & (0/1,1) & ${}^1S_0,{}^3P_0,{}^3P_1,{}^3P_2-{}^3F_2,{}^1D_2\ldots$ \\
       c2ct0 & N3LO & (0,1) & $^1S_0,{}^1D_2\ldots$ \\
       c2ct1 & N3LO & (1,1) & ${}^3P_0,{}^3P_1,{}^3P_2-{}^3F_2\ldots$ \\ \hline
       \multicolumn{4}{c}{Tensor} \\ \hline
       c2t0  & NLO & (1,0)   & ${}^3S_1-{}^3D_1,{}^3D_2,{}^3D_3-{}^3G_3\ldots$ \\
       c2t1  & NLO & (1,1)   & ${}^3P_0,{}^3P_1,{}^3P_2-{}^3F_2\ldots$ \\
       c4t0  & N3LO & (1,0)   & ${}^3S_1-{}^3D_1,{}^3D_2,{}^3D_3-{}^3G_3\ldots$ \\
       c4t1  & N3LO & (1,1)   & ${}^3P_0,{}^3P_1,{}^3P_2-{}^3F_2\ldots$ \\ \hline
       \multicolumn{4}{c}{Tensor Isotensor} \\ \hline
       c2tt  & N3LO & (1,1) & ${}^3P_0,{}^3P_1,{}^3P_2-{}^3F_2\ldots$ \\ \hline
       \multicolumn{4}{c}{Spin-Orbit} \\ \hline
       c2b   & NLO & (1,0/1) & ${}^3S_1-{}^3D_1,{}^3P_0,{}^3P_1,{}^3P_2-{}^3F_2,{}^3D_2\ldots$ \\ 
       c4b0  & N3LO & (1,0)   & ${}^3S_1-{}^3D_1,{}^3D_2,{}^3D_3-{}^3G_3\ldots$ \\
       c4b1  & N3LO & (1,1)   & ${}^3P_0,{}^3P_1,{}^3P_2-{}^3F_2\ldots$ \\ \hline
       \multicolumn{4}{c}{Spin-Orbit Isotensor} \\ \hline
       c2bt  & N3LO & (1,1) & ${}^3P_0,{}^3P_1,{}^3P_2-{}^3F_2\ldots$ \\ \hline
       \multicolumn{4}{c}{$(\text{Spin-Orbit})^2$} \\ \hline
       c4bb  & N3LO & (1,0/1) & ${}^3S_1-{}^3D_1,{}^3P_0,{}^3P_1,{}^3P_2-{}^3F_2,{}^3D_2\ldots$ \\ \hline
       \multicolumn{4}{c}{$(\text{Orbital Angular Momentum})^2$} \\ \hline
       c4q0  & N3LO & (0,0/1) & ${}^1P_1,{}^1D_2\ldots$ \\
       c4q1  & N3LO & (1,0/1) & ${}^3S_1-{}^3D_1,{}^3P_0,{}^3P_1,{}^3P_2-{}^3F_2,{}^3D_2\ldots$ \\ 
   \end{tabular}
   \caption{Table of LECs ordered by operators. We list the order of the LEC, $(S,T)$ channel pair, and partial waves acted upon by the given LEC. }
   \label{tab:LECs}
   \end{ruledtabular}
\end{table}

\subsection{Regularization and coordinate space interaction}
\label{subsec:coord}
In order to use the contact interactions, divergences must be removed, which is accomplished through using a regularizer. For our purposes, we choose a Gaussian cutoff of the form~\cite{Piarulli:2014bda, Piarulli:2016vel, Schiavilla:2021dun}
\begin{equation}\label{eq:cutoffp}
    \widetilde{C}_{R_s}(q) = e^{-R_s^2 q^2/4},
\end{equation}
which removes the divergent high-momentum content of the interaction. In coordinate space, Eq.~(\ref{eq:cutoffp}) leads to a local regulator given by
\begin{equation}
    C_{R_s}(r) = \frac{1}{\pi^{3/2}R_s^3}e^{-(r/R_s)^2}.
\end{equation}
The set of cutoffs we have investigated are $R_s = \lbrace 1.5, 2.0, 2.5  \rbrace$ fm, corresponding to typical momentum-scale cutoffs $\Lambda_s=2/R_s$ from about 263 MeV down to 158 MeV.

Note that in Ref.~\cite{Schiavilla:2021dun}, the regulator used is dependent on the isospin channel of the operator, with different cutoffs for the $T=1$ and $T=0$ channels. This was necessary in their fitting procedure due to the requirement of $c0c0$ and $c0c1$ to reproduce the large singlet and triplet scattering lengths, which induces strong correlations. Thus, two different cutoffs for $T=0$ and $T=1$ were used to reduce these correlations. In this work, we opt to employ a single cutoff for the two isospin channels. We do this due to the employment of a Bayesian framework for the model calibration, which will act to change the constraints from Ref.~\cite{Schiavilla:2021dun}. These details are discussed in Sec.~\ref{sec:methods}. Moreover, the choice of two cutoffs in the isospin channels is not a convenient option for performing calculations of larger nuclei with, for instance, the auxiliary field diffusion Monte Carlo method~\cite{Schmidt:1999lik,Carlson:2014vla,Gandolfi:2020pbj}, as it introduces isospin-dependent spin-orbit terms at lower orders in the chiral expansion, which are challenging to treat with this type of method~\cite{Lonardoni:2017hgs,Lonardoni:2018nob,Piarulli:2019pfq,Schiavilla:2021dun}.

\section{Methods}
\label{sec:methods}

\subsection{Bayesian Parameter Estimation Framework}
\label{subsec:bayes}

In this work, the development of contact interactions is closely integrated with the implementation of a Bayesian model calibration approach developed by the BUQEYE Collaboration~\cite{Schindler:2008fh,Wesolowski:2018lzj, Melendez:2019izc, Wesolowski:2021cni}. Our primary goal is the estimation of the LECs that parameterize the $NN$ interaction discussed in Sec.~\ref{sec:int}.
Recently, Detmold et al.~\cite{Detmold:2023lwn} have made estimations of $\slashed{\pi}$EFT LECs from Lattice QCD finite-volume spectra, though these estimates currently apply to pion masses significantly higher than the physical value. Here, we follow the more traditional approach of estimating the LECs by calibrating to $NN$ scattering data.
We begin by defining a probability distribution for the LECs, of which we have 2(11)\{26\} at LO(NLO)\{N3LO\}, that are collected into a parameter vector we denote as $\mathbf{a}$. We apply Bayes' theorem to derive their posterior distribution given the data $\mathbf{y}$, i.e., observables (which is a separate space from the parameter space and thus generally of different dimension), and external information $I$,
\begin{equation}\label{eq:lec_posterior}
{\rm pr}(\mathbf{a}|\mathbf{y},I) \propto \mathrm{pr}(\mathbf{y} \vert \mathbf{a}, I) \mathrm{pr}(\mathbf{a}\vert I),
\end{equation}
where $\mathrm{pr}(\mathbf{a}|\mathbf{y},I)$ is the posterior distribution we seek. This distribution is derived by combining the likelihood, $\mathrm{pr}(\mathbf{y} \vert \mathbf{a}, I)$, with the prior, $\mathrm{pr}(\mathbf{a}\vert I)$, which encodes any prior knowledge about $\mathbf{a}$.
 
This approach readily aligns with more conventional model calibration methods, such as weighted least-squares. In a simplified, uncorrelated model space, the Bayesian likelihood uses the traditional goodness of fit measure, the venerable $\chi^2$,
\begin{equation}\label{eq:lec_likelihood}
\mathrm{pr}(\mathbf{y} \vert \mathbf{a}, I) \propto e^{-\chi^2/2}.
\end{equation}
Furthermore, adopting a uniform prior and seeking the point that maximizes ${\rm pr}(\mathbf{a}|\mathbf{y},I)$, the Bayesian framework corresponds precisely to a least-squares approach where minimizing the $\chi^2$ maximizes both the likelihood and the posterior.

However, outside of this context, the traditional and Bayesian approaches quickly diverge. In the least-squares approach, we attempt to calibrate our model to minimize the $\chi^2$. Conversely, in the Bayesian approach, we seek to find the total distribution for the posterior such that we are presented with a probability for a given parameter set. One may ask why we choose a Bayesian approach over a least-squares approach. The first obvious reason is that we can begin to extract parametric uncertainty for the model from the posterior. Further motivations for a Bayesian approach are more veiled but yield much more powerful consequences. We can begin to see this if we write Eq.~(\ref{eq:lec_likelihood}) in a more illuminating manner,
\begin{equation}\label{eq:cor_likelihood}
    \mathrm{pr}(\mathbf{y} \vert \mathbf{a}, I) \propto e^{-\left(\mathbf{y}_{\mathrm{exp}} - \mathbf{y}_\mathrm{th}(\mathbf{a})\right)^{\rm T} \Sigma^{-1} \left(\mathbf{y}_{\mathrm{exp}} - \mathbf{y}_\mathrm{th}(\mathbf{a})\right)}.
\end{equation}
In this form, the likelihood begins to direct us away from the standard $\chi^2$ goodness of fit. For example, if we add correlations in the covariance matrix present in Eq.~(\ref{eq:cor_likelihood}), we already diverge from the $\chi^2$ indicator. Further, we can construct a covariance that takes uncertainties besides only uncertainties on $\mathbf{y}$ into account. This idea is the foundation to our Bayesian framework, where our goal is to encode EFT truncation uncertainties in the likelihood. This is laid out in Sec.~\ref{subsec:ther_errors}.

\subsection{Theoretical Errors}
\label{subsec:ther_errors}

In our effort to completely characterize the uncertainties associated with the $\slashed{\pi}$EFT model, it becomes necessary to incorporate a measure of theoretical errors in the model calibration. To accomplish such a task, we must produce an estimate of the error associated with our model of choice. Our implementation of such a theoretical error follows along the treatment of Ref.~\cite{Wesolowski:2018lzj}, where a treatment of EFT truncation error is added. The exact propagation of the inclusion of such errors can be complicated; however, we can treat a simple linear model for some observable $y_i$:
\begin{equation}
    y_{{\rm exp},i}(x) = y_{{\rm th},i}(x;\mathbf{a}) + \delta y_{{\rm th},i}(x;\mathbf{a}) + \delta y_{{\rm exp},i}(x).
\end{equation}
Thus, our treatment of total error in the calibration of our EFT models enters the LEC likelihood as an addition of the experimental discrepancy, $\delta y_{\rm exp}$, and the theoretical discrepancy, $\delta y_{\rm th}(\mathbf{a})$. Further, to simplify the error propagation, we assume that $\delta y_{\rm exp}$ and $\delta y_{\rm th}$ are normally distributed. This assumption allows for the simple addition of the experimental and theoretical variances in quadrature. 

To arrive at a form for $\delta y_{\rm th}$, we make a connection to the perturbation expansion of the EFT~\cite{Furnstahl:2015rha}. That is to say that we assume $y_{\rm th}$ can be written as a power series 
\begin{equation} \label{eq:series_expansion}
    y_{\rm th}(x;\mathbf{a}) = y_{\rm ref}(x)\sum_{n=0}^\infty c_{2n}(x;\mathbf{a}) Q^{2n}(x),
\end{equation}
where $Q$ is the EFT expansion parameter given in Eq.~(\ref{eq:eft_expansion_parameter}). It is important to note here that the assumed series expansion in Eq.~(\ref{eq:series_expansion}) is an expansion in $Q^2$, which arises due to the specific choice of Weinberg power counting having contact interactions only at even powers of $Q$. We also include a reference scale, $y_{\rm ref}$, for the series expansion of the theoretical calculations to set the overall magnitude and dimension of the quantities. All of the scales in the expansion are encoded in $y_{\rm ref}$ and $Q$, which thus means that the expansion coefficients,  $c_n$, are dimensionless and roughly natural, i.e., of order 1. We emphasize that the expansion coefficients are \textit{not} the LECs of the EFT, but instead arise from the order-by-order correction of the EFT.

The series in Eq.~(\ref{eq:series_expansion}) is treated up to some order $n=k$ such that the expansion in Eq.~(\ref{eq:series_expansion}) matches the working order in $Q$ of the EFT expansion. Thus, this becomes
\begin{equation}
    y_{\rm{th}}^{(k)}(x;\mathbf{a}) = y_{\rm ref}(x)\sum_{n=0}^k c_{2n}(x;\mathbf{a}) Q^{2n}(x),
\end{equation}
where the neglected part of the series is
\begin{equation} \label{eq:series_trunc}
    \delta y_{\rm{th}}^{(k)}(x;\mathbf{a}) = y_{\rm ref}(x)\sum_{n=k+1}^\infty c_{2n}(x;\mathbf{a}) Q^{2n}(x).
\end{equation}
We can thus associate Eq.~(\ref{eq:series_trunc}) with the uncertainty that arises from truncating the perturbative expansion of the EFT. Upon inspection, we can see that the EFT expansion parameter dictates that this is a convergent series. In fact, under the assumption that the truncation error is uncorrelated across orders, this can be recognized as a geometric series in $Q^2$, which gives rise to an analytic form for the truncation error:
\begin{equation}\label{eq:th_err}
    \delta y_{\rm th}^{(k)}(x;\mathbf{a}) = \frac{y_{\rm ref} \,  \cbar \, Q^{2(k+1)}}{\sqrt{1-Q^4}},
\end{equation}
where $\cbar^2$ is the marginal variance of the populations of $c_n$s, $c_n | \cbar \sim \mathcal{N}(0,\cbar^2)$. Given that the expansion coefficients govern the order-by-order convergence scheme of the Eq.~(\ref{eq:series_expansion}), we expect these to be roughly natural and thus for $\cbar$ to be natural as well.

Thus, armed with the uncertainty in the form of Eq.~(\ref{eq:th_err}) for some given observable $y_i$, we can begin to formulate a covariance of the truncation uncertainty for a given set of observables. Since we have assumed Gaussian random variables for the theoretical discrepancy, we can utilize the closure of Gaussian variables under addition and multiplication to give the covariance~\cite{Wesolowski:2018lzj}. This allows us to find the theoretical covariance due to the truncation as 
\begin{equation}\label{eq:cor_err}
    \left(\Sigma_{\rm th} \right)^{\rm corr.}_{ij} = \frac{\left(y_{{\rm ref},i} \, \cbar \, Q_i^{2(k+1)}\right)\left(y_{{\rm ref},j} \, \cbar \, Q_j^{2(k+1)}\right)}{1-Q_i^2Q_j^2} r(x_i,x_j;l),
\end{equation}
where indices $i$ and $j$ index observables $y_i$ and $y_j$. Further, we have included a function, $r(x_i,x_j;l)$, which enters into the covariance to smooth and reduce the strength of correlations between observables. This function's exact form depends on the problem being investigated; our choice is discussed in Sec.~\ref{subsec:parm_est}. The need to introduce such dampening and smoothing functions for the correlations is both physically and numerically motivated. First, we expect data that are separated by a large distance in the physical input space not to have strong correlations. Second, if we allow full or even strong correlations to exist, the covariance can have a tendency to become singular or ill-conditioned, preventing the inversion of the matrix, as is necessary in Eq.~(\ref{eq:cor_likelihood}).

We can, however, also choose to completely break the correlations in the covariance by enforcing a strictly diagonal covariance, 
\begin{equation}\label{eq:uncor_err}
    \left(\Sigma_{\rm th} \right)^{\rm uncorr.}_{ij} = \left(\Sigma_{\rm th} \right)^{\rm corr.}_{ij}\delta_{ij}.
\end{equation}
The obvious action of breaking the correlation is the enforcement that each piece of data adds the same amount of information in the calibration. This can be a useful treatment if we are unsure of the correlation structure of the data or if there is no physical motivation for correlation in the errors that define the discrepancy between theory and data. But we note here that this should be handled carefully. If we have a large amount of data that are close in the physical input space and are thus correlated, the uncorrelated treatment will over-constrain the model in such a case due to the over-accounting of effective data. Regardless of the motivation for the treatment of an uncorrelated model, by the use of Eq.~(\ref{eq:uncor_err}), we are able to include some measure of truncation uncertainty in the model calibration. However, when compared to a truncation model that includes correlations, we expect an uncorrelated model to offer stronger constraints.

Once we have a covariance matrix to represent the theoretical uncertainty, either correlated or uncorrelated, we construct a total covariance matrix via the addition of the experimental covariance and the theoretical covariance under quadrature, 
\begin{equation}\label{eq:tot_cov}
    \Sigma_{ij} = \left( \Sigma_{\rm exp} \right)_{ij} \delta_{ij} + \left( \Sigma_{\rm th} \right)_{ij}.
\end{equation}
In general, experimental uncertainty is reported without correlations, so the contribution from the experimental error is entered as a diagonal element in the total covariance. Thus, the covariance is diagonal for uncorrelated theory error models, whereas theoretical correlations are the only contributions to off-diagonal elements. Finally, armed with the covariance, we can introduce the correlated form of the $\chi^2$, known as the Mahalnobis distance,
\begin{equation}
    d_M(\mathbf{a}) = (\mathbf{y}_{\rm exp} - \mathbf{y}_{\rm th})^T \Sigma^{-1} (\mathbf{y}_{\rm exp} - \mathbf{y}_{\rm th}),
\end{equation}
which becomes our measure that enters into the likelihood $\rm{pr}(\bf{y}|\bf{a},I)$, which is the exact form that we find in the exponent of Eq.~(\ref{eq:cor_likelihood}). Due to the nature of including additional errors, $d_M$ will behave slightly differently than the traditional $\chi^2$. If we examine how $d_M$ behaves with only diagonal contributions from the truncation error to the covariance matrix, then $d_M$ will be less than if only experimental errors are accounted for. This can act to reduce much of the constraint placed by the experimental data. If we allow off-diagonal elements in the covariance, the effective number of data is reduced, and the correlations introduced can strengthen some of the constraints. Thus, we may naively expect that $d_M^{\rm uncorr.} \leq d_M^{\rm corr.}$.

However, our construction of the total posterior remains incomplete due to the introduction of $\Lambda_b$ and $\cbar$ in $\Sigma$, which both have their own pdfs that we must account for. 

\subsection{Estimation of \texorpdfstring{$\Lambda_b$ and $\cbar$}{Lambdab and cbar}}\label{subsec:lamb_and_cbar}

With the inclusion of $\Lambda_b$ and $\cbar$ in the model calibration, the total posterior that we must now estimate becomes~\cite{Wesolowski:2021cni}
\begin{align}\label{eq:total_post}
    \mathrm{pr}(\mathbf{a},\cbar^2,\Lambda_b|\mathbf{y},I) \propto \,
    &{\rm pr}(\mathbf{y} | \mathbf{a},\cbar^2,\Lambda_b,I)  \mathrm{pr}(\mathbf{a}|I) \times \nonumber \\
    &\mathrm{pr}(\cbar^2|\Lambda_b,\mathbf{a},I) \mathrm{pr}(\Lambda_b|\mathbf{a},I).
\end{align}
We have already discussed the formulation of the first term, the likelihood, at given values of $\bar{c}^2$ and $\Lambda_b$. The second term is a prior on the LECs, that we need to choose ourselves, based on an interpretation of what it means for these parameters to be natural. 

That leaves the posteriors for $\cbar^2$ and $\Lambda_b$  still to be explored. Inside each of these posteriors we will find further priors, which we again must choose ourselves. We note here that the overall posterior in Eq.~(\ref{eq:total_post}) may be somewhat affected by the choice of priors in our Bayesian framework. While some may see the inclusion of priors as a weakness of Bayesian analysis, this, in fact, offers a significant advantage over a standard least-squares minimization approach as we must make explicit statements about any assumptions. Any disagreement with an analysis performed in a Bayesian framework can easily be done under different assumptions and, thus, different priors. We begin the construction of Eq.~(\ref{eq:total_post}).

First, for $\cbar^2$, as recognized in Sec.~\ref{subsec:ther_errors}, this is the marginal variance of the expansion coefficients. So we can follow the normal practice, where unknown variances have priors given by a scaled inverse chi-square distribution~\cite{Wesolowski:2021cni},
\begin{equation}
    \rm{pr}(\cbar^2|I) \sim \chi^{-2}[\nu_0,\tau_0^2],
\end{equation}
where initial hyperparameters $\nu_0$, the degrees of freedom, and $\tau_0$, the scaling parameter, are left as a choice. Our choice of prior is highly motivated by two points. First, the posterior is trivial to find, as the scaled inverse chi-square distribution is given by conjugate prior-posterior pairs, i.e., the posterior is also a scaled inverse chi-square distribution:
\begin{equation}\label{eq:cbar_post}
    \mathrm{pr}(\cbar^2|\mathbf{a},\Lambda_b,I) \sim \chi^{-2}[\nu,\tau^2(\mathbf{a},\Lambda_b)].
\end{equation}
Second, in the limit of a large number of data, by choosing the marginal variance $\cbar$ to be modeled by the scaled inverse chi-square distribution, we can make a direct link between the scale $\tau^2$ and $\cbar$. This simply means that the most probable value for $\cbar^2$ is $\tau^2$ for a given set of $(\mathbf{a},\Lambda_b)$.

With this, all that we must do is find the updated formula for our hyperparameters. For $\nu$, this is rather straightforward, as this is given by the degrees of freedom of the distribution, which is to say the number of data we have,
\begin{equation}\label{eq:nu}
    \nu = \nu_0 + N_{\rm obs} n_c.
\end{equation}
Here we have the prior choice, $\nu_0$, the total number of observables that we calibrate to, $N_{\mathrm obs}$, and the number of orders of the EFT expansion that we include in the order-by-order analysis, $n_c$.

For the scale parameter, we must first find the expansion coefficients, which can be seen in Eq.~(\ref{eq:series_expansion}) to be 
\begin{equation}\label{eq:expansion_parameters}
    c_{n,i}({\bf a},\Lambda_b) = \frac{y_n^{(i)}({\bf a}_{(n)}) - y_{n-1}^{(i)}({\bf a}_{(n-1)})}{y_{{\rm ref},i}Q_i^{2n}},
\end{equation}
where $n$ indexes the order of the interaction and $i$ indexes discrete observables. From the coefficients, we then construct the scale parameter as 
\begin{equation}\label{eq:tau}
    \nu \tau^2(\mathbf{a},\Lambda_b) = \nu_0 \tau_0^2 + \sum_{n,i}c_{n,i}^2(\mathbf{a},\Lambda_b).
\end{equation}
Looking closer at this, we can identify that for $N_{\rm obs} n_c \gg \nu_0$, the scale parameter simply becomes the root mean square of the expansion coefficients. This, in turn, is related to the scaled order-by-order correction of the EFT. With Eqs.~(\ref{eq:cbar_post}), (\ref{eq:nu}), and (\ref{eq:tau}), we can evaluate the posterior for any $\cbar$.

For the posterior for $\Lambda_b$, we are also fortunate that we can find an analytic form. This follows from Ref.~\cite{Melendez:2019izc}, where a rigorous derivation can be found. For brevity, we simply give the form as 
\begin{equation}\label{eq:lamb_post}
    \mathrm{pr}(\Lambda_b|\mathbf{a},I) \propto \frac{\mathrm{pr}(\Lambda_b|I)}{\tau^\nu \prod_{n,i}Q_i^{2n}}.
\end{equation}
Here we make note that the product in the denominator runs over both all of the orders $n$ present in the order-by-order analysis, as well as over all of the observables $i$. It is important to note that the form given in Eq.~(\ref{eq:lamb_post}) is unnormalized. For the $\Lambda_b$ posterior, unlike all of the other posteriors presented so far, it must be normalized in order to accurately calculate the true value of the posterior in any calibration process. This is due to the implicit dependence on $\mathbf{a}$ in $\tau$ that appears in the $c_n$'s, which means the normalization is also dependent on $\mathbf{a}$. This does not pose a substantial problem numerically, only that the normalization must be performed. 

The choice of prior $\mathrm{pr}(\Lambda_b|I)$ can generally be any choice. However, there is one property of the prior that must be maintained. We must encode that $\mathrm{pr}(\Lambda_b \leq p_{\mathrm{max}}|I) = 0$, where $p_{\mathrm{max}}$ is the maximum momentum used in the calibration of the EFT model. This is a trivial characteristic of the prior for $\Lambda_b$, as if this is not maintained, Eq.~(\ref{eq:series_expansion}) will be divergent for some observables. This also naturally constrains the expansion parameter to be less than 1, which is also an expectation built into the EFT.

We have now quantified every part of the posterior in Eq.~(\ref{eq:total_post}). All that remains now is the explicit choice of prior for the LECs $\mathbf{a}$, for $\Lambda_b$, the choice of $r(x,x';l)$ that enters into Eq.~(\ref{eq:cor_err}), choices for $\nu_0$ and $\tau_0$, and a means to estimate the total posterior. This discussion follows in the next section.

\subsection{Parameter Estimation for \texorpdfstring{\pionlessEFT}{Pionless EFT}}
\label{subsec:parm_est}

To estimate the posterior given in Eq.~(\ref{eq:total_post}), we turn to Markov Chain Monte Carlo (MCMC) methods. The use of MCMC allows us to draw effective samples from the distribution and generate statistics to quantify the unknown distribution. Our choice for the MCMC sampler is the off-the-shelf Python package \texttt{emcee}~\cite{Foreman-Mackey:2012any}. From \texttt{emcee}, we utilize the implemented parallelized stretch move to draw many simultaneous samples, allowing for quick exploration of the parameter space. The choice of an off-the-shelf package is a simple one, as we only need to supply a means to evaluate the posterior to the MCMC algorithm. 

The data used in the calibration is found in neutron-proton ($np$) and proton-proton ($pp$) scattering observables in the Granada database~\cite{Perez:2013jpa, Perez:2013oba, Perez:2014yla}.
The database contains scattering data taken from 1950 to 2003, which were then compiled into a 3$\sigma$ self-consistent database. In the arrangement of the database, there are $N$ sets of data corresponding to different experiments. Each data set contains
measurements at fixed lab energy, $E_{\mathrm{lab}}$, and different scattering angles, $\theta$. However, a few observables are measured at different $E_{\mathrm{lab}}$ and fixed $\theta$, like, for example, total cross sections since their measurement does not involve the scattering angle ($\theta=0$). The data blocks have an associated normalization factor and systematic uncertainty, which arises in the consistent construction of the database. See Ref.~\cite{Piarulli:2014bda} for more details.

For the construction of the EFT, the momentum scale that we use in Eq.~(\ref{eq:eft_expansion_parameter}) is the center of mass momentum, $p_{\mathrm{c.m.}}$, where kinematics gives
\begin{equation}\label{eq:cm_momentum}
    p_{\rm{c.m.}} = \sqrt{E_{\mathrm{lab}} \mu},
\end{equation}
where $\mu$ is the reduced mass of the scattering system. The center of mass momentum is encoded in a vector, along with the scattering angle, $\mathbf{x} = (p_{\mathrm{c.m}},\theta)$. Further, we include the deuteron binding energy and $nn$ scattering length, ${}^1a_{nn} = -18.9$ fm~\cite{Chen:2008zzj}, as data in the EFT calibration. 

As Eqs.~(\ref{eq:cor_err}), (\ref{eq:expansion_parameters}), and  (\ref{eq:lamb_post}) require the EFT expansion parameter given by Eq.~(\ref{eq:eft_expansion_parameter}), we require the use of soft momentum scales. While it is possible to estimate these quantities as part of the parameter estimation process, we choose to fix the soft scales at relevant physical scales for the type of scattering system involved. The natural scales for these interactions are given by the deuteron binding for $np$, whereas the Coulomb-modified $pp$ scattering length is a natural choice for $pp$ scattering. Thus, the soft scales that we chose are given by 
\begin{equation}\label{eq:soft_scales}
    p_{\mathrm{soft}} =\begin{cases}
        p_D \approx 45\text{ MeV/c}, & \text{for $np$ scattering} \\
        1/{}^1a_{pp} \approx 25\text{ MeV,} & \text{for $pp$ scattering}.
    \end{cases}
\end{equation}
As with all things done in a Bayesian fashion, these soft scales are, in fact, parameters that we could estimate. However, we have not done so here as this falls outside of the scope of the LEC and truncation estimation we aim to implement in this paper. Thus, viewed in a Bayesian fashion, we have chosen the priors for the soft scales to be delta-functions centered at our choice. 

Continuing on with defining our priors, we look at the first prior that we introduced, which was $\mathrm{pr}(\mathbf{a}|I)$. For the choice of this prior, we lean on a physical motivation, as well as one that arises computationally. Physically, the LECs that arise out of $\chi$EFT and $\slashed{\pi}$EFT are natural, i.e., of order 1 when expressed in the appropriate units. Thus, one might be inclined to choose a prior of the form $\mathrm{pr}(\mathbf{a}|I) \sim \mathcal{N}(0,\sigma^2)$, with uncorrelated variances for each of the LECs. This choice is a rather natural one, as this is a much more informative prior than a uniform distribution. However, in the case of our $\slashed{\pi}$EFT, the choice for the prior becomes less straightforward. This is due to two things:
\begin{enumerate}
    \item The projected LECs lack the naturalness of a basis of non-projected constants. 
    \item The magnitude of the LECs pick up a cutoff dependence as the cutoff stiffens. 
\end{enumerate}
To navigate these two effects, we thus chose our prior with the information from other computational findings in the model calibration.

We still chose a normal distribution for our exact choice of prior, but we chose the multi-dimensional mean to be non-zero. To find the mean of the Gaussian, we can turn to a different data set to inform the prior. In particular, we can use the Granada phase shifts \cite{Perez:2013jpa,Perez:2013oba,Perez:2014yla}. Whereas it is common practice to use phase shifts as pseudo or surrogate data in model calibration~\cite{Gezerlis:2014zia,Wesolowski:2018lzj,Somasundaram:2023sup}, we can treat the phase shift data set as a lower fidelity data set as there is model dependence in their extraction from scattering data. Further, one naively expects there to be large correlations in energy for phase shifts, reducing the number of effective data when calibrating to them. This leads to configurations of LECs with large uncertainty when performing a Bayesian calibration. However, we gain valuable information by calibration to the lower fidelity phase-shits. If we perform a calibration to the phase shift data, we can find the configuration with the highest posterior, $\mathbf{a}_{\mathrm{p.s.}}.$ This maximum posterior configuration we use to encode the prior for calibration to the scattering observables, $\mathrm{pr}(\mathbf{a}|I) \sim \mathcal{N}(\mathbf{a}_{\mathrm{p.s.}},\sigma^2)$. This approach of informing the model by lower fidelity simulations is akin to history matching~\cite{Bower:2010,Vernon:2014} where we identify regions of non-implausibly. Further, a connection can also be made to sequential design approaches~\cite{Surer:2024}, where calibration to low-fidelity models can be used to identify areas of interest in high-fidelity modeling. Such approaches have already been studied in the context of EFT model calibration \cite{Hu:2021trw,Thim:2023fnl} and may be necessary for future Bayesian model calibration of EFTs.

The choice for the variance is left completely as a choice. For calibrating our $\slashed{\pi}$EFT, we chose a variance of $\sigma^2 = 10^2$ in the relevant units for each LEC. We chose quite a large variance as we still want a slightly uninformative prior since we retain quite a bit of uncertainty about the exact magnitude of the LECs. If a highly informative, i.e., low variance, prior is chosen, then the posterior can easily become dominated by the prior. Further, even if the prior is not highly constraining, a highly peaked Gaussian for the prior can break any correlations in the LEC posterior. Thus, it is highly desirable in the case of the calibration of these particular $\slashed{\pi}$EFT to maintain a rather uncertain prior, which we achieve with our specific choice. Our prior serves to bound the parameter space to a physical region but does not highly constrain it otherwise. 

For $\Lambda_b$, we are also required to supply a prior for the analysis. For $\chi$EFT that have $\pi$ degrees of freedom, previous works have estimated that the breakdown scale should be $\approx 600\text{ MeV}$~\cite{Furnstahl:2015rha,Melendez:2017phj,Reinert:2017usi}. This breakdown scale corresponds roughly to the masses of the light $\rho$ and $\omega$ mesons, which are integrated out into the contact interactions present in $\chi$EFT. For $\slashed{\pi}$EFT, the $\pi$ is further integrated out of the theory, leaving only contact interactions in the theory. Following the same logic as $\chi$EFT, the naive expectation is $\Lambda_b \approx m_\pi$ in $\slashed{\pi}$EFT. Previous work has been done by Lepage to extract breakdown scales~\cite{Lepage:1997cs}, which falls in line with this expectation. Given the naive expectation and its verification by Lepage, one might be tempted to encode a strong prior around $m_\pi$. However, as of yet, no statistical inference has been performed to estimate the $\Lambda_b$. From this missing piece of information, we choose instead a weak prior of the form
\begin{equation}
    \mathrm{pr}(\Lambda_b|I) \sim \mathcal{N}(\mu = 500, \sigma = 1000)
\end{equation} 
which is truncated on the low end at $p_{\mathrm{max}}$ and 700 MeV at the high end. A Gaussian prior was chosen to add more information than a simple uniform prior, with a bias in between the naive estimations for $\Lambda_b$ for $\pi$-full and $\slashed{\pi}$ interactions. We choose a large variance to maintain a weakly informative prior. Further, we restrict $\Lambda_b < 700$ MeV for computational efficiency, as we do not expect higher breakdowns than this scale.

The priors for $\cbar$ are a less complicated choice due to using a standard choice of an inverse chi-square distribution. As such, we don't need to further motivate a distribution for the prior. However, we still must provide a choice for the prior degree of freedom, $\nu_0$, and scaling parameter, $\tau_0$. For these, we follow in the footsteps of \cite{Wesolowski:2021cni}, and we choose $\nu_0 = \tau_0 = 1.5$. For this previous work, there was minor sensitivity to the choice of priors of the hyperparameters as the number of data entered in the model calibration was quite small. Thus, the analysis for $\cbar$ and likewise $\Lambda_b$ could produce dependence on the prior choice for $\nu_0$ and $\tau_0$. However, in the present analysis, the Granada database contains a large amount of data. Due to this fact, the degrees of freedom and scaling parameter of the inverse chi-square distribution are dominated by the number of data and order-by-order convergence of the models. Therefore any choice for these hyperparameters that take a natural scale for the $\chi^{-2}$ prior will have no effect on the conjugate posterior as the hyperparameters are updated.

Finally, the remaining choice that must be made is $r(\mathbf{x},\mathbf{x}';\mathbf{l})$ in Eq.~(\ref{eq:cor_err}). To smooth our correlations, we use a simple exponential kernel of the form
\begin{equation}\label{eq:kernal}
    r(\mathbf{x},\mathbf{x}';\mathbf{l}) = e^{\frac{|p_{\mathrm{c.m.}}-p'_{\mathrm{c.m.}}|}{l_p}}e^{\frac{|\theta-\theta'|}{l_\theta}}\delta_{\rm {type,type'}},
\end{equation}
where $l_p$ and $l_\theta$ are correlation lengths for the center of mass momentum and scattering angle, respectively. In practice, the correlation lengths can be calibrated in some way, such as by training a Gaussian process \cite{Svensson:2023twt}. Or one could do a similar analysis to that done for $\cbar$ and $\Lambda_b$ to estimate probability distributions. For our analysis, we have opted to fix the correlation lengths at $l_p = 0.3$ MeV and $l_\theta = 20^\circ$. These values were chosen as they maintain correlations for data at similar energies and scattering angles while reducing the strength of the correlations for data separated in phase space. We have not performed an analysis to optimize these parameters in any way. Instead, as this is a study in the implementation of Bayesian calibration methods to estimate the LECs and truncation uncertainties, we leave this for further study.

Our choices of priors are ultimately motivated to be informative to some extent but not to be as constraining as the likelihood in the model calibration. This is due to the fact that we remain rather uncertain of the \textit{a priori} constraints that we can place on the quantities we seek to estimate. While there exists a substantial amount of physical motivation for the priors, as discussed, due to the exact nature of the models being investigated, we cannot place a large amount of information in the physical constraints. The uncertainties that arise due to cutoff effects and the sensitivity of specific data to the model make placing strong constraints via a prior a difficult task. However, the flexibility allows anyone to redo the model calibration with their own assumptions and priors if they desire to do so. 

\section{Analysis and Results}
\label{sec:results}
\subsection{LEC Estimation without Truncation}
\label{subsec:no_trunc}

The starting point of our analysis to estimate the LECs of \pionlessEFT\ is through treatment without accounting for any model discrepancies. Such a treatment is through the implementation of at Eq.~(\ref{eq:lec_posterior}) with a likelihood given by Eq.~(\ref{eq:lec_likelihood}). The exact form for the $\chi^2$ that enters into the likelihood can be found in Ref.~\cite{Piarulli:2014bda}.

\begin{figure}[ht]
    \begin{center}
    \includegraphics[width=\linewidth]{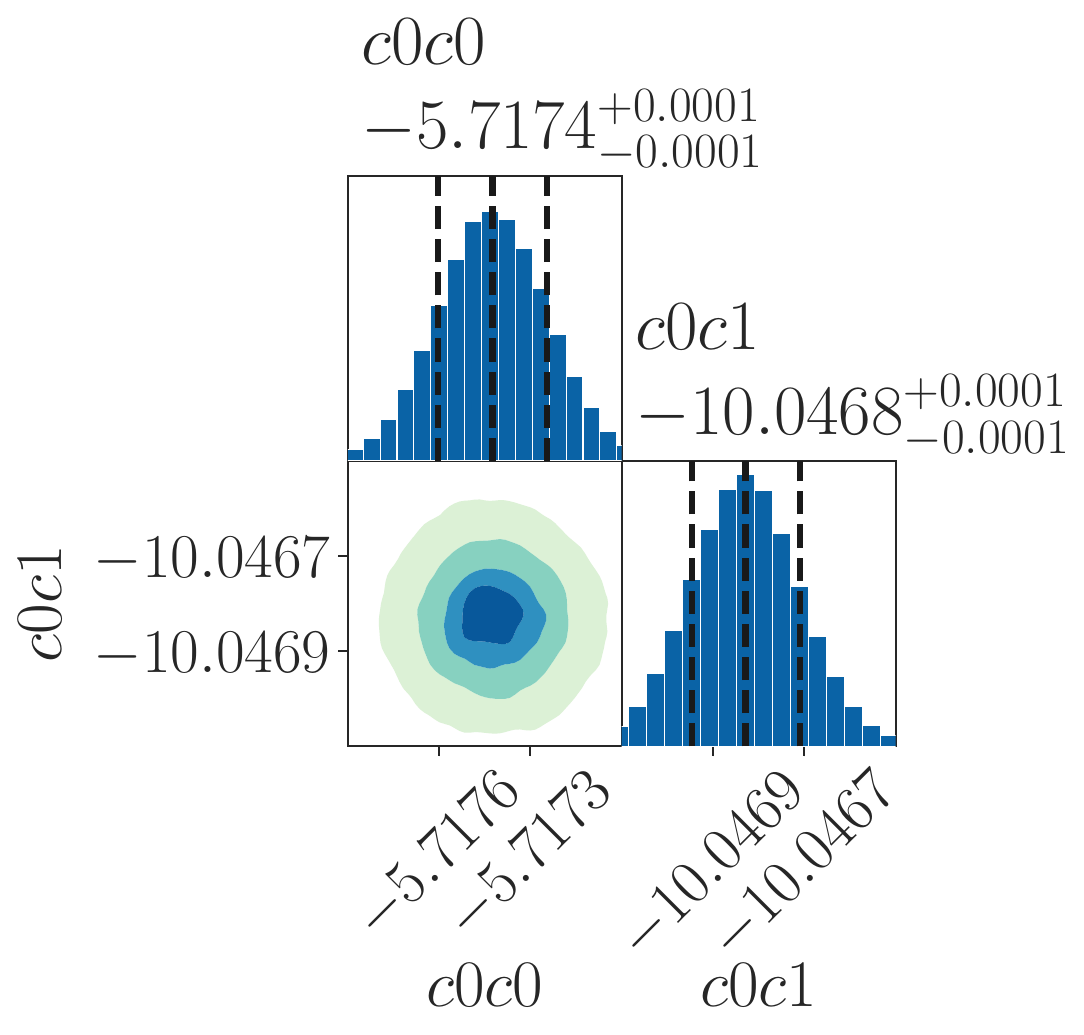}
    \end{center}
    \caption{Two-dimensional joint and one-dimensional marginal distributions for LO \pionlessEFT\ LECs with $R_s = 2.0$ fm with LECs presented in the projected $S=0,1$ basis. The vertical lines denote the 16th, 50th, and 84th percentiles. Contours are given at the 12th, 39th, 68th, and 86th percentiles.}
    \label{fig:2.0.lo.no.trunc.corner}
\end{figure}

Following conventional expectations of fitting a low-energy EFT to $NN$ scattering data, our initial step involved the optimization of the LO pionless EFT models. This optimization was carried out utilizing $np$ scattering data up to $E_{\rm lab}=0.5$ MeV, comprising of 48 total cross section measurements~\cite{Perez:2013jpa,Perez:2013oba,Perez:2014yla}, the deuteron binding energy, and $nn$ scattering length. Fig.~\ref{fig:2.0.lo.no.trunc.corner} showcases the joint and marginal distributions for the projected LO LECs in the channels $(S,T)=(0,1)$ and $(S,T)=(1,0)$, $c0c0$ and $c0c1$ respectively, using a cutoff of $R_s = 2.0$ fm.

From this figure, we can see that the projected LO interactions have LECs that are independently distributed and symmetrically centered around their median values. When examining the overall magnitudes of the LECs, one could be concerned about size of the $c0c1$ LEC, as it is drifting away from naturalness. However, one must remember that these LECs are not the standard $C_S$ and $C_T$ but are linear combinations of them; refer to Sec.~\ref{subsec:Projected Basis}. When transforming back to the standard basis, the LECs take natural values. The interested reader can find the transformed figure of Fig. \ref{fig:2.0.lo.no.trunc.corner}, along with similar figures for other cutoffs, in a repository that we have created for this paper~\cite{beft}.

At NLO, the inclusion of a more intricate operator structure allows for a significant extension of the energy range considered in the calibration process. We have considered $np$ and $pp$ data up to $E_{\rm lab}$=15 MeV, the deuteron binding energy, and $nn$ scattering length.  The data for the NLO calibrations is thus a larger data set, with 746 data points, including 213 total cross sections, 15 longitudinal and transverse total cross sections, 425 differential cross sections, 88 polarizations, and 5 asymmetries~\cite{Perez:2013jpa,Perez:2013oba,Perez:2014yla}.

\begin{figure*}[ht]
    \begin{center}
        \includegraphics[width=\linewidth]{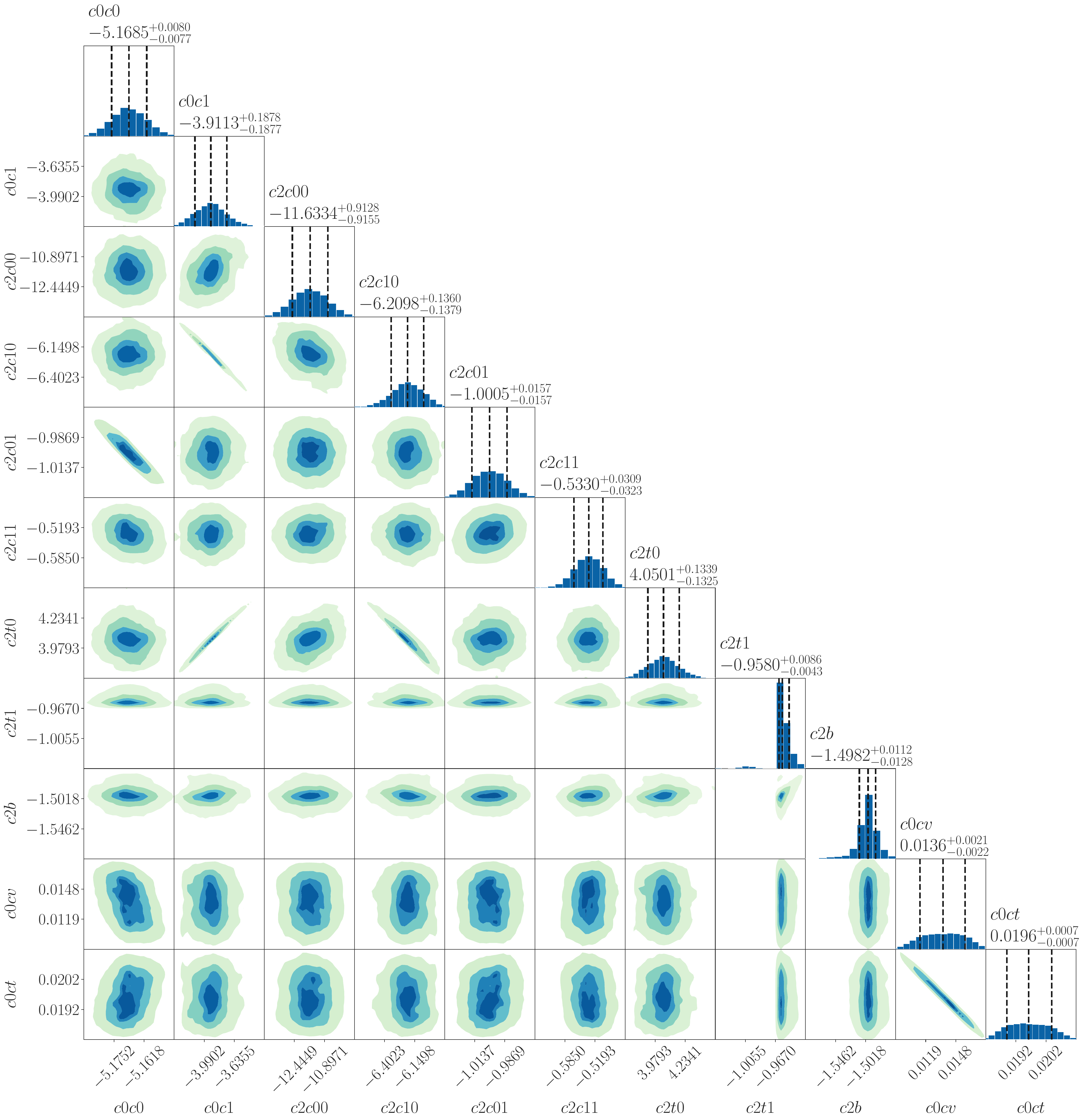}
    \end{center}
    \caption{Joint and marginal distributions for NLO LECs with $R_s = 2.0$ fm. The LECs are presented in a projected $S=0,1$ and $T=0,1$ basis. See Fig.~\ref{fig:2.0.lo.no.trunc.corner} for figure notation.}
    \label{fig:2.0.nlo.no.trunc.corner}
\end{figure*}

A figure of the joint and marginal distributions for NLO LECs projected in the $S = 0, 1$ and $T = 0, 1$ basis using cutoff $R_s = 2.0$ fm is given in Fig. \ref{fig:2.0.nlo.no.trunc.corner}.

As we include more parameters, a pattern emerges showing the correlations among parameters operating within the same channel. For the central potential, we see correlations among the LO and NLO LECs in the $(S,T)=(0,1)$ ($c0c0$ and $c2c01$), and $(S,T)=(1,0)$ channels ($c0c1$ and $c2c10$), respectively. Further, the $(S,T)=(1,0)$ central LECs ($c0c0$ and $c2c10$) share a correlation with the tensor in the $T=0$ ($c2t0$). There also exists a correlation between the $T=1$ tensor ($c2t1$), and spin-orbit parameter ($c2b$). With the $T=1$ tensor LEC ($c2t1$), we also notice a distinct cut in the distribution below the peak of the distribution. This possibly arises due to constraints placed by observables sensitive to $p$-waves, particularly in the coupled-channel, which restricts the region for $c2t1$. Further, we can notice that $c2t1$ and $c2b$ are highly peaked, indicating a shift away from a normal distribution. 

Likewise, we see non-normal higher-order moments for the charge-dependent parameters ($c0cv$ and $c0ct$) as these distributions are quite broad. These two parameters also exhibit a strong anti-correlation, indicating that they act to break charge independence in an opposing manner. This is not unexpected as they operate in the same channels, as noted in Tab.~\ref{tab:LECs}. Similar distributions and correlation structures for $R_s = 1.5$ fm can be found in the repository~\cite{beft}, along with distributions presented in the standard basis.

Once we have estimations for the LECs, we can begin to look at how the model performs in calculating observables. We can do this in two ways. First, we can use the standard maximum a posteriori (MAP) estimator, which is the ``best'' model, as in standard practice. Or, we can generate a posterior predictive distribution (PPD), effectively propagating the parametric uncertainty to observables, which is found bound marginalizing over the LECs,
\begin{equation}\label{eq:ppd}
    \mathrm{pr}(\mathbf{y}_{\mathrm{th}}| \mathbf{y},I) = \int d\mathbf{a} \, \mathrm{pr}(\mathbf{y}_{\mathrm{th}} | \mathbf{a}, I) \mathrm{pr}(\mathbf{a}|\mathbf{y}, I),
\end{equation}
where $\mathrm{pr}(\mathbf{y}_{\mathrm{th}} | \mathbf{a}, I)$ is deterministic for a given value of $\mathbf{a}$ and is thus a delta-function. This can effectively be found by calculating $y_{\mathrm{th}}$ for LEC sets sampled from $\mathrm{pr}(\mathbf{a}|\mathbf{y}, I)$. For the interactions described by the posteriors in Figs.~\ref{fig:2.0.lo.no.trunc.corner} and \ref{fig:2.0.nlo.no.trunc.corner}, we assessed applications to phase shifts and deuteron properties to both verify that the MCMC sampling yields valid models and to quantify the parametric uncertainty that the posteriors generate for observables. In this exercise, we found that the parameter estimation uncertainties on these observables are on the order of less than 1\%.

These small parametric uncertainties are what we expect from these models. Since we have only included experimental discrepancies in the calibration, the constraints on the LECs are quite strong. Such constraints greatly limit the effect that any parametric uncertainties in the LECs have on the uncertainties in observables. In order to generate larger parametric variances, we must reduce the constraints on the LECs, which can be done via the inclusion of model discrepancies. Specifically, we expect that the truncation errors inherent in the EFTs will emerge as the primary source of uncertainty, and we thus want to investigate the truncation contributions.

\subsection{Convergence of LECs with Truncation Errors}
\label{subsec:convergence}
To begin our investigation of the contribution of truncation errors in observable estimation, we first begin with their role in model calibration. In particular, it is of great interest to look at the stability of the LEC estimation as a function of $E_{\mathrm{lab}}^{(\mathrm{max})}$. For LEC estimation done without any accounting of model uncertainties, one should naively expect two things. First, the LECs may vary as a function of $E_{\mathrm{lab}}^{(\mathrm{max})}$. This arises due to the fact that all data is only weighted by the experimental uncertainties and thus are treated roughly equally since experimental uncertainties tend to have the same relative scales. So, the inclusion of more data should constantly change constraints. Second, LECs that are estimated without model discrepancies should take on different values than those done with model discrepancies. This is because, again, without model discrepancy, all data is given roughly the same relative weight in the calibration. Both of these points are key conclusions from Ref.~\cite{Wesolowski:2018lzj}, where calibrations were performed using phase shifts. Here, it is prudent to carry out a similar analysis since we use scattering data directly. 

To this end, we implement Eq.~(\ref{eq:tot_cov}) to include model uncertainties in the calibration process. We can include the model discrepancy in two different ways. First, we can treat an uncorrelated model, which has the inclusion of $\delta_{ij}$ in Eq.~(\ref{eq:cor_err}), and second, we can follow the correlation construction laid out in Sec.~\ref{subsec:parm_est} with Eq.~(\ref{eq:kernal}).
At this stage, we must still choose a value for $\Lambda_b$ and $\cbar$ in the analysis, as we are interested solely in the effect of the truncation error on the constraints on the LECs. For this study, we fix $\Lambda_b = m_\pi$ and $\cbar = 1.5$. These choices are naive expectations, where we are assuming that Eq.~(\ref{eq:series_expansion}) is well-behaved. It is important to note here that these truncation errors are sensitive to $\Lambda_b$. For $np$ systems below $p_{\mathrm{soft}}=p_d$, we can find that increasing $\Lambda_b=m_\pi$ by 20\% yields $\approx$100\% increase in the truncation error at NLO. 

\begin{figure*}[h!tb]
    \begin{center}
        \includegraphics[width=0.97\linewidth]{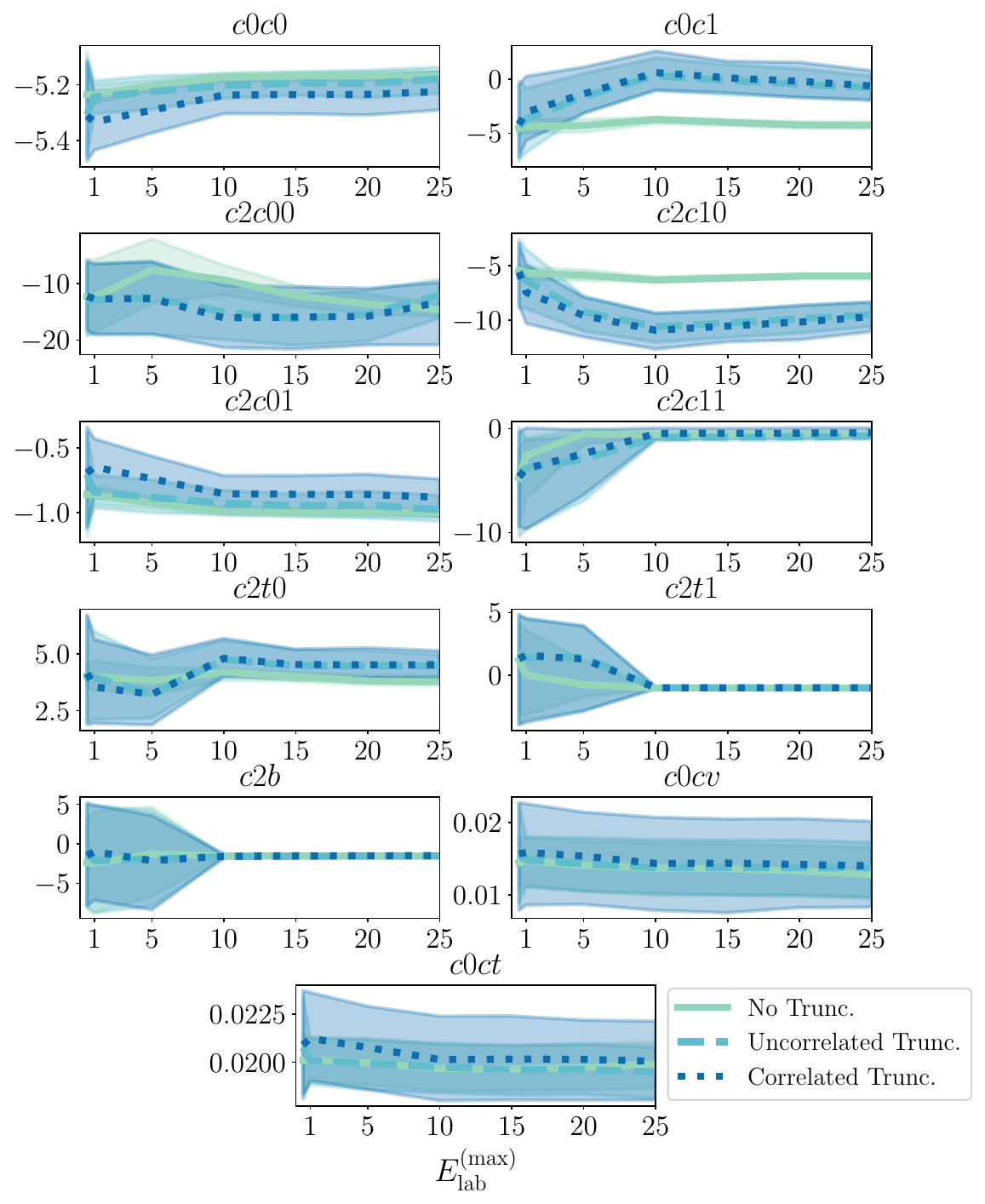}
    \end{center}
    \caption{The convergence of the NLO LECs for a 2.0 fm cutoff with different truncation error models included in the calibration as a function $E_{\mathrm{lab}}^{(\mathrm{max})}$. The solid teal, dashed cyan, and dotted blue lines correspond to calibrations done with zero, uncorrelated, and correlated truncation errors, respectively, with the shaded region corresponding to the 95\% credibility level. $\Lambda_b = m_\pi$ and $\cbar = 1.5$ were fixed.}\label{fig:Wesolowski_2_0_fm}
\end{figure*}

We can examine these two uncertainty models across a range of maximum lab energy used in the calibration, which is shown in Fig. \ref{fig:Wesolowski_2_0_fm} for a 2.0 fm cutoff at NLO. For calibrations that do not include any accounting for the model discrepancy, we can find that the LECs generally have very tight constraints. This leads to little variability in the LECs as higher energy scattering data is included. Once we account for the model discrepancy in the calibration, we find that these behaviors vanish.

With the inclusion of model discrepancies, we see much looser constraints on the LECs, signified by their larger variance. The reduction in constraint is also apparent as we vary the maximum lab energy of the scattering data included in the calibration. As more data is included, the LECs undergo larger variations due to the reduction of the influence of low-energy data in the calibration. This trend of LEC variation continues up to 10 MeV maximum lab energy, where the LECs generally plateau to a value that is stable with the inclusion of higher energy data.

When seeking the cause of this ``phase change'' in the behavior of the LECs, it is particularly insightful to examine the LECs active in the $(S,T)=(0,0)$ and $(S,T)=(1,1)$ channels, where contributions from $p$-waves are most significant. In particular, the central ($c2c11$), the tensor ($c2t1$), and spin-orbit ($c2b$) LECs have a general trend of becoming highly constrained at $E_{\rm{lab}}^{\rm{(max)}}=10$ MeV.
This could be attributed to the inclusion of a larger number of data that has a strong dependence on $p$-wave contributions. At 5 MeV, there are a total of 388 scattering data comprised of 133 total cross sections, 4 longitudinal and transverse total cross sections, and 251 differential cross sections, whereas by the time we reach 10 MeV the database includes an additional 50 total cross sections, 6 longitudinal and transverse total cross sections, 123 differential cross sections, 42 polarizations, and 1 asymmetry. The polarized cross sections add sensitivity to the ${}^1P_1$, ${}^3P_0$, and ${}^3P_1$ phase shifts~\cite{Wilburn:1995zz, Walston:2001yx} generating additional constraints, whereas the polarizations and asymmetry place constraints on ${}^3P$ channels~\cite{Barker:1982zz, Slobodrian:1967, Hutton:1975, Weisel:1992zz, Obermeyer:1980bew}. Thus, these additional data between 5 and 10 MeV, contribute to the dramatic convergence of $(S,T)=(1,1)$ LECs $c2c11$, $c2t1$, and $c2b$, and the added minor convergence of $(S,T)=(0,0)$ LEC $c2c00$.

Further, the inclusion of more data alters the constraints on $(S,T)=(0,1)$ and $(S,T)=(1,0)$ channel central ($c0c0$, $c0c1$, $c2c10$, and $c2c01$) and tensor ($c2t0$) LECs, which likewise exhibit a weaker convergence between $E_{\rm{lab}}^{\rm{(max)}}=[5,10]$ MeV. It is intriguing that this behavior coincides with the convergence of the $(S,T)=(0,0)$ and $(S,T)=(1,1)$ LECs. This could be attributed to three effects. First, the increase in polarized cross section data places additional constraints on ${}^1S_0$, ${}^3S_1$, ${}^3D_1$, and $\epsilon_1$ mixing angle. Second, at 5 MeV $c2c00$, $c2c11$, $c2t1$, and $c2b$ are relatively unconstrained due to the lack of $p$-wave content. This requires the constraints placed on $c0c0$, $c0c1$, $c2c10$, $c2c01$, and $c2t0$ to be altered to make accurate estimations of the polarized cross sections. Once more $p$-wave constraints are added, then $c0c0$, $c0c1$, $c2c10$, $c2c01$, and $c2t0$ no longer have the ``artificial" balancing constraints.

The final contribution to the convergence of the $(S,T)=(0,1)$ and $(S,T)=(1,0)$ LECs comes from $d$-wave contributions. For $E_{\rm{lab}}^{\rm{(max)}}=10$ MeV and above, $d$-waves start to pick up meaningful strength. Thus, they are no longer negligible as they contribute more to cross sections. Further, as we work with a regularized interaction, the regularization itself contributes to artifacts. These artifacts arise in $d$-waves and higher channels and thus offer no change in constraint in $s$- and $p$-waves. For our regularization scheme, the $d$-wave artifacts contribute a non-inconsequential amount to the overall strength of the $d-$wave content. Thus, these artifacts cannot be overlooked in their contribution to the convergence pattern, which alters the constraints on $c0c0$, $c0c1$, $c2c10$, $c2c01$, and $c2t0$ in a non-trivial way as more data is added.

With the analysis of how the LECs behave as we change the maximum lab energy, we can see that there is little need to include scattering data greater than 10 MeV lab energy in the calibration. Doing so introduces no extra constraints on the LECs and only makes the calibration more computationally complex. However, we must remember that this analysis follows for a single fixed value for $\Lambda_b$. This behavior could change if the true breakdown scale is different, as this changes the constraints of the model discrepancy.

For any parameter estimation that we choose to perform, we must balance the computational cost of the scattering calculations while maintaining a large enough value of $E_{\rm{lab}}^{\rm{(max)}}$ to have enough constraints to place on the LECs. This becomes even more imperative when we include estimations of $\Lambda_b$ and $\cbar$ via the full posterior given in Eq.~(\ref{eq:total_post}).

\subsection{Full Bayesian Parameter Estimation}
\label{subsec:model_calibration}

In this section, we finalize our analysis by calculating the complete posterior as defined in Eq.~(\ref{eq:total_post}). The individual posteriors for $\cbar$ and $\Lambda_b$ are specified in Eqs.~(\ref{eq:cbar_post}) and (\ref{eq:lamb_post}), respectively.
Through the inclusion of the posteriors for $\cbar$ and $\Lambda_b$ and, we can fully quantify aspects of the EFT models ranging from parameterization to truncation error. 

\begin{figure*}
    \centering
    \includegraphics[width=\linewidth]{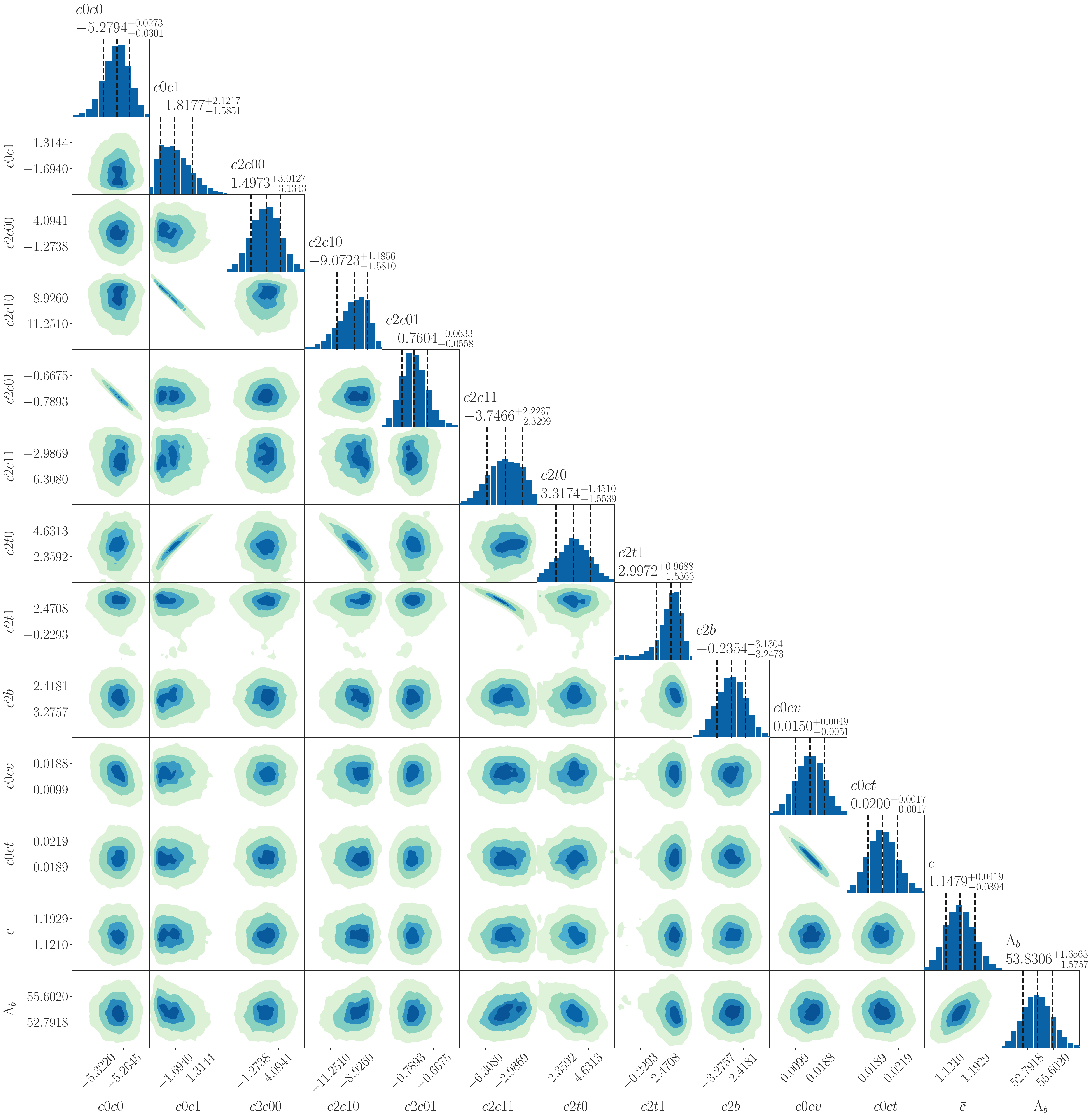}
    \caption{Posterior distributions for $\slashed{\pi}_{\text{WB}}$NLO$_{\text{Full}}$-2.0 LECs. See Fig.~\ref{fig:2.0.lo.no.trunc.corner} for figure notation.}
    \label{fig:NLO_2.0_fm_corner}
\end{figure*}

To calculate Eq.~(\ref{eq:total_post}) for sub-leading orders, we must begin our analysis at LO, estimating Eq.~(\ref{eq:lec_posterior}) with some fixed value of $\Lambda_b$ and $\cbar$. Following the analysis in Sec.~\ref{subsec:convergence}, a good, naive starting point for the model calibration is to use $\Lambda_b = m_\pi$, $\cbar=1.5$ and use data up to $E_{\rm lab} = 10$ MeV, which can be adjusted once breakdown estimates are made. From the estimation of Eq.~(\ref{eq:lec_posterior}) for LO, we can then estimate Eq.~(\ref{eq:total_post}) for NLO. For the calculation of the coefficients given in Eq.~(\ref{eq:expansion_parameters}), we use MAP estimators for the $NN$ observables. That is to say, for the LO observables in Eq.~(\ref{eq:expansion_parameters}), we use $y_0^{(i)}(\mathbf{a}_0^{\mathrm{MAP}})$. We label interactions done in the order-by-order calibration as $\slashed{\pi}_{\text{WB}}$(NLO/N3LO)$_{\text{Full}}$ and present the joint and marginalized pdfs given by Eq.~(\ref{eq:total_post}) for $\slashed{\pi}_{\text{WB}}$NLO$_{\text{Full}}$-2.0 in Fig.~\ref{fig:NLO_2.0_fm_corner}. Figures for $R_s=[2.5,1.5]$ fm can be found in the associated repository~\cite{beft}. 

From this corner plot, we can start to see the correlation structure of the interaction in the various channels in which each LEC acts. For instance, the central LO coefficient in the $(S,T)=(0,1)$ channel, $c0c0$, exhibits a strong correlation with the corresponding central NLO LEC, $c2c01$. Similarly, in the $(S,T)=(1,0)$ channel, the central LO coefficient, $c0c1$, is strongly correlated not only with the central NLO LEC, $c2c10$ but also with the tensor NLO LEC of the same channel, $c2t0$. Additionally, the tensor NLO LEC, $c2t0$, and the central NLO LEC, $c2c10$, in the $(S,T)=(1,0)$ channel are strongly correlated. The correlations also extend to the $(S,T)=(1,1)$ NLO central and tensor LECs, $c2c11$ and $c2t1$, respectively.  And, finally, a significant correlation is observed among the charge-dependent LECs, $c0cv$ and $c0ct$. This is very similar to the correlation structure that we observe in Fig.~\ref{fig:2.0.nlo.no.trunc.corner} for the calibration performed without any inclusion of truncation uncertainties. 

The novel behavior for the order-by-order analysis lies in the parameters that govern the truncation error. As noted in Sec.~\ref{subsec:lamb_and_cbar}, the value of $\cbar$ for large collections of data is given by the r.m.s. error of the series expansion coefficients from Eq.~(\ref{eq:expansion_parameters}). As we expect these coefficients to be natural for a well-behaved model, we should expect the population variance to be roughly natural as well, which is what we observe for $\cbar$ in Fig. \ref{fig:NLO_2.0_fm_corner}, and likewise for the other cutoffs, which can be seen the repository. However, we can notice a large change in the naive estimation of $\Lambda_b = m_\pi$ and the order-by-order estimations of $\Lambda_b \in [52.2,55.4]$ MeV at the 68\% credibility level for $R_s=2.0$ fm at NLO. This estimation for $\Lambda_b$ is applicable only to the Weinberg power counting that we have employed, and we stress that other power counting schemes may yield different estimations. Similar values are found for the other cutoffs, which can be seen in the repository~\cite{beft}. Not only is this estimation far lower than one might expect, but the preceding analysis in Sec.~\ref{subsec:convergence} for the LEC extraction might follow a different trend. However, while the convergence may no longer follow as expected, the information gained in the previous exercise still provides insight into the extraction at the current breakdown.

First, we must have $p_{\rm{c.m.}}/\Lambda_b \leq 1$, which gives the maximum lab energy that can be used in the analysis. From Eq.~(\ref{eq:cm_momentum}), we can find that 
\begin{equation}\label{eq:E_max}
    E_{\rm{max}} = \frac{\Lambda_b^2}{\mu}.
\end{equation}
For a lower bound of $\Lambda_b=50$ MeV, we find $E_{\rm{max}}\approx 5$ MeV, which we choose as the maximum energy for $E_{\rm{lab}}$. However, from Sec.~\ref{subsec:convergence}, we found that the bulk of the data that constrains $p$-waves enters between 5 and 10 MeV lab energies. This indicates that there is little data to constrain LECs that operate in the central and tensor $(S,T)=(0,0)$ and $(S,T)=(1,1)$ channels as well as the spin-orbit. This is further indicated in Fig. \ref{fig:Wesolowski_2_0_fm} as there is little change in these LECs between calibrations done with 1 and 5 MeV max lab energies. From this observation, we can see that these LECs contribute little to the estimates of low-energy observables, where the LECs mostly likely have constraints from the requirement of reproducing physical observables. This observation does not change from Fig.~\ref{fig:Wesolowski_2_0_fm} to the full calibrations shown in Fig.~\ref{fig:NLO_2.0_fm_corner}. In these distributions, we find poorly constrained $(S,T)=(0,0)$ and $(S,T)=(1,1)$ LECs, where the 1$\sigma$ level can be as large as the median value of the LEC. Thus, there is a strong indication that these LECs are superfluous for describing the interaction.

Furthermore, if we look at the data in the Granada database, we have 4 data up to 5 MeV lab energy that would constrain LECs acting in the $p$-wave. These are the polarized total cross sections, $\Delta \sigma_T$~\cite{Wilburn:1995zz} and $\Delta \sigma_L$ \cite{Walston:2001yx}. These 4 data have large associated experimental uncertainties, generating a weak experimental constraint. Further, the largest driving factor in the weak constraints comes from the truncation uncertainty. These data lie near the edge of the kinematically allowed momentum domain, each having a large associated EFT expansion parameter, $Q$. Due to these weak constraints, we will explore interactions that neglect $p$-wave physics, which we discuss in Sec.~\ref{subsec:s_wave_interactions}.

With NLO interactions, we can carry on with the order-by-order analysis of the EFT to calibrate models at N3LO for the $\slashed{\pi}_{\text{WB}}$N3LO$_{\text{Full}}$ set of models. Due to the kinematic constraints at NLO, the N3LO models must also be done up to lab energies of 5 MeV. Figures for the $\slashed{\pi}_{\text{WB}}$N3LO$_{\text{Full}}$ posteriors can be found in the repository~\cite{beft}. As with the NLO interactions, we expect weak constraints on the $(S,T)=(0,0)$ and $(S,T)=(1,1)$ channels as well as the spin-orbit LECs due to the relative lack of data in these channels. Further, we can notice an extension to the correlation behavior present in the $\slashed{\pi}_{\text{WB}}$NLO$_{\text{Full}}$ models to the $\slashed{\pi}_{\text{WB}}$N3LO$_{\text{Full}}$ models.

\begin{figure}
    \centering
    \includegraphics[width=\linewidth]{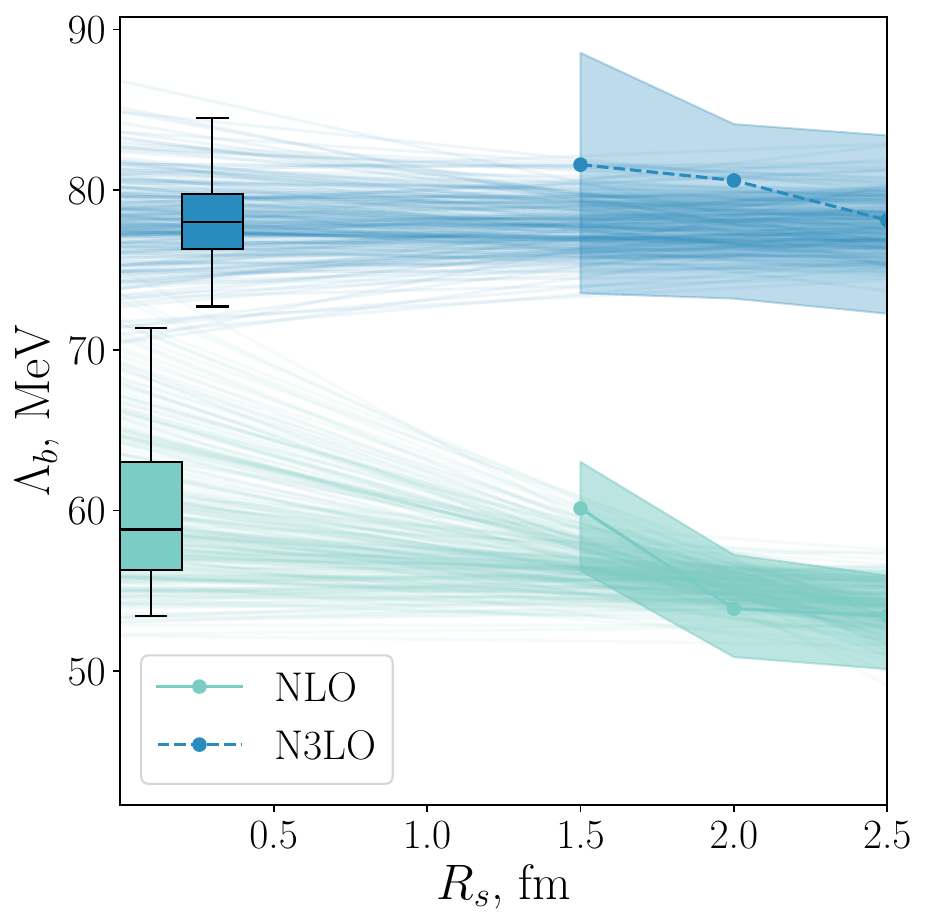}
    \caption{The estimations of $\Lambda_b$ for $\slashed{\pi}_{\text{WB}}$NLO$_{\text{Full}}$ and $\slashed{\pi}_{\text{WB}}$N3LO$_{\text{Full}}$ models given by Eq.~(\ref{eq:lamb_post}). The error bands on the estimates represent the 95\% credibility level for the posteriors. Linear fit estimates found with MCMC are also plotted. The distributions for $\Lambda_b$ extrapolated to $R_s=0$ are represented with box-and-whisker plots. The boxes denote the standard 1st, 2nd, and 3rd quartiles definitions for boxplots. For the whiskers, we adapt the 95\% credibility level for the distributions.}
    \label{fig:breakdown_trends}
\end{figure}

However, with the $\slashed{\pi}_{\text{WB}}$N3LO$_{\text{Full}}$ estimates, one can immediately notice a large discrepancy in the estimation of $\Lambda_b$. Estimates of the breakdown vs.\ the cutoff are shown in Fig.~\ref{fig:breakdown_trends}.  When taking our analysis to N3LO, we find a jump in $\Lambda_b \in [75.4 ,80.9]$ MeV at the 68\% credibility level. Further, the breakdown estimates at N3LO are less constrained, as indicated by the larger credibility level. In the next subsection, we discuss a possible physical origin for these shifts in the preferred value of $\Lambda_b$ from NLO to N3LO.

\subsection{Order Dependence of Breakdown}
\label{subsec:breakdown_order_dependence}
As noted in Sec.~\ref{subsec:model_calibration}, we find  NLO and N3LO estimates of $\Lambda_b$ that do not agree at the 95\% credibility level. As $\Lambda_b$ is physically linked to the energy scale above which degrees of freedom are integrated out and replaced by contact interactions, the order-by-order estimation of $\Lambda_b$ in our analysis is rather unexpected. For any EFT, as we increase the order of the interaction, we expect a refinement to the description of the interaction but not a change in the physical scale of the processes involved. Thus, this discrepancy may be indicative of an external source of uncertainty that we have not included in the model calibration. 

To address this issue, we first investigate the regularization scheme that we employ for the $\slashed{\pi}$EFT. As with any finite regularization, regulator artifacts are always introduced in the interaction. This can be seen in the different estimates for the LECs, $\cbar$, and $\Lambda_b$ across the different cutoffs, where the figures are found in the repository. However, as $R_s \rightarrow 0$ ($\Lambda_s \rightarrow \infty$),  these cutoff-dependent artifacts are expected to diminish, leading to a more consistent breakdown scale, $\Lambda_b$, across different orders of the theory. Therefore, it becomes revealing to examine estimates of $\Lambda_b$ as $R_s \rightarrow 0$. It must be noted here that we cannot perform calculations of the scattering data in a Weinberg power counting scheme at small values of $R_s$ due to the Wigner bound. Due to this, we perform a simple linear extrapolation of the breakdown scales to $R_s = 0$.

In order to estimate what the breakdown would be at $R_s = 0$ fm, we must have some form for the trend of the breakdown as we reduce the cutoff. For the sake of simplicity, we can assume a linear trend. Further, we can perform the linear estimation with MCMC to generate posteriors for the slope and intercept to get a more rigorous estimate. In the MCMC sampling, we assume that the breakdowns are independently distributed as non-central $t$-distributions at each cutoff, which we characterize by fitting. We then take the product of these $t$-distributions to be the likelihood. 

For the prior on the intercept, we assume a uniform prior over $\Lambda_b(R_s = 0\text{ fm}) \in [0,m_\pi]$. For the slope, the naive uninformative prior would be a uniform prior over a range of slopes. However, this prior is biased towards larger slope angles due to their higher density when sampled uniformly. Instead, a prior on the slope that is uniform over the angle maximizes the entropy, leading to a maximally non-informative prior. The prior that accomplishes this is 
\begin{equation}
    \mathrm{pr}(m|I) \propto (1+m^2)^{-3/2},
\end{equation}
where here $m$ denotes the slope. With our likelihood and priors, we can then create a credibility level for the breakdown extrapolated to $R_s = 0$ fm for $\slashed{\pi}_{\text{WB}}$N3LO$_{\text{Full}}$ models. Figure~\ref{fig:breakdown_trends} shows 200 sampled linear fits from the MCMC for the two orders. We show the extrapolated $\Lambda_b(R_s = 0)$ via boxplots, where the whiskers give the 95\% credibility level. From these posteriors, we find the 95\% credibility levels for the extrapolated breakdowns, which are $\Lambda_b^{\rm{(NLO)}}(R_s = 0)\in [53.414,71.374]$ MeV and $\Lambda_b^{\rm{(N3LO)}}(R_s = 0) \in [72.723,84.498]$ MeV for NLO and N3LO respectively. Thus, after extrapolating to remove regulator artifacts, we find that there is no agreement at the 95\% level and cannot say that the discrepancy between $\Lambda_b$ between NLO and N3LO is attributable in large part to regulator artifacts. 

The other origin of the discrepancy may arise due to the series expansion in Eq.~(\ref{eq:series_expansion}). In our formulation of the order-by-order problem, we chose an expansion in $Q^2$, as this matches the natural order of the EFT in the Wienberg power counting scheme. However, in making this choice, we may introduce a systematic underestimation of the truncation errors in each order, as we do not treat the missing orders. For our order-by-order analysis, we treat NLO and N3LO truncation errors at $Q^4$ and $Q^6$ while the trailing orders are $Q^3$ and $Q^5$, respectively. This effectively underestimates the truncation error by $Q^2$ at each order. 

Due to this systematic underestimation of the truncation error in the treatment of the truncated order, the missing error must be accounted for in some way. This accounting in the analysis can only be handled by $\Lambda_b$, as $\cbar$ plays a much smaller role in the truncation estimate. Thus, in order to balance the underestimated modeled truncation error, there must be a suppression in $\Lambda_b$. Through this suppression, the truncation error is able to maintain the magnitude that is informed by the data. This motivates a future examination of a regulated $\slashed{\pi}$EFT in a standard power counting scheme~\cite{Kaplan:1998tg,Kaplan:1998we} that has interactions present at all orders of $Q$. 

\subsection{$s$-wave-dominant Interactions}
\label{subsec:s_wave_interactions}

As discussed in Sec.~\ref{subsec:model_calibration}, the kinematics of the estimated $\Lambda_b\approx 50$ MeV at NLO requires the calibration of our EFTs to data below 5 MeV in the lab frame. This requirement eliminates data that would constrain central and tensor LECs acting in $p-$ and higher-odd partial waves, namely $c2c00$, $c2c11$, $c2t1$, $c4c00$, $c4c11$, and $c4t1$, and spin-orbit and orbital momentum squared LECs, namely $c2b$, $c4b0$, $c4b1$, $c4bb$, $c4q0$, and $c4q1$. To this end, we have performed the calibration of models setting these LECs to zero so that the interaction is predominantly one that encodes $s$-wave physics. This aligns well with the expectation of such a low-energy theory: low-energy physics should generally be described with an $s$-wave interaction.

\begin{figure*}
    \centering
    \includegraphics[width=\linewidth]{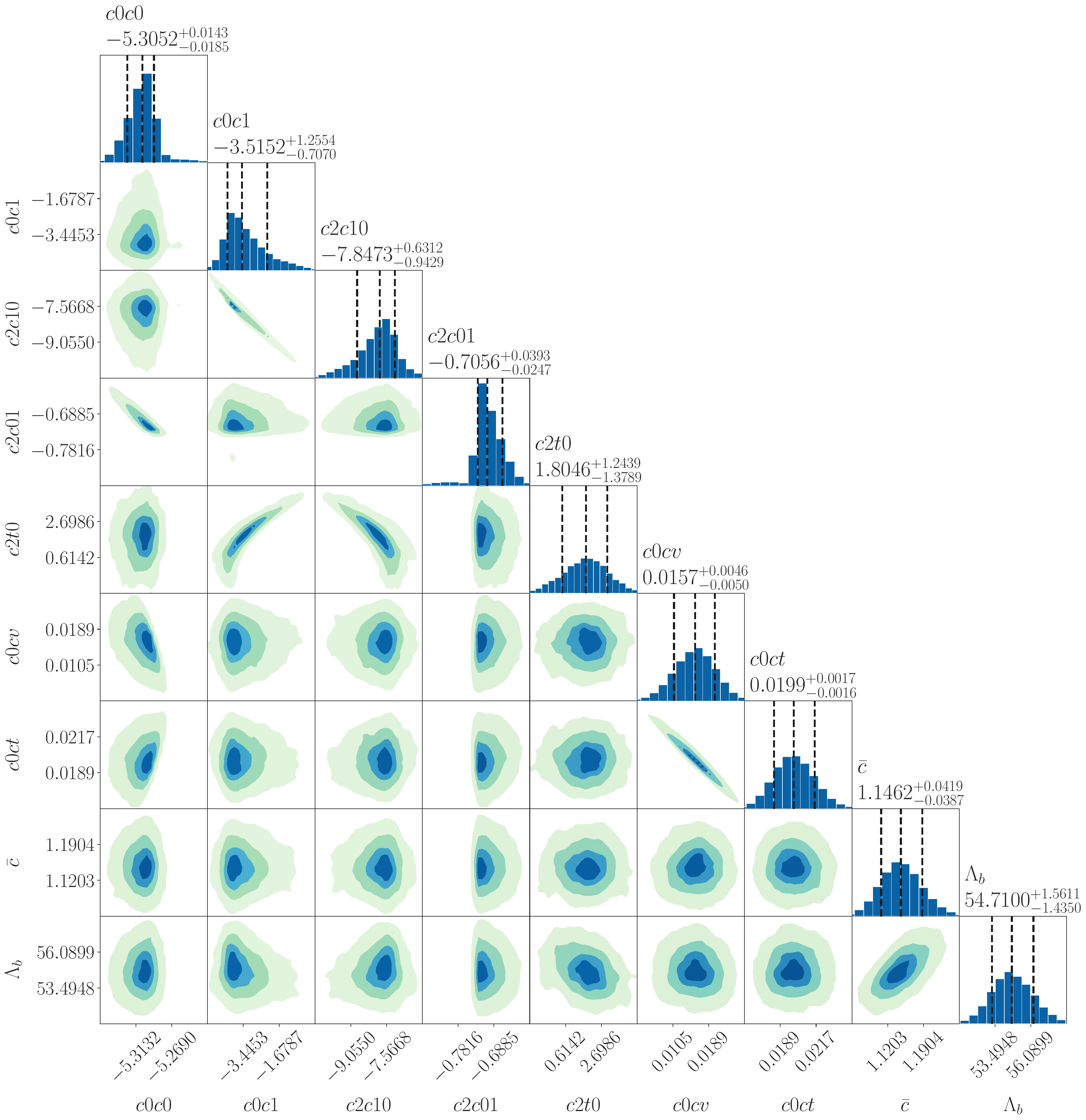}
    \caption{Posterior distributions for $\slashed{\pi}_{\text{WB}}$NLO$_{\text{Red.}}$-2.0 LECs. See Fig.~\ref{fig:2.0.lo.no.trunc.corner} for figure notation.}
    \label{fig:2.0.nlo.s.wave.corner}
\end{figure*}

\begin{figure*}
    \centering
    \includegraphics[width=\linewidth]{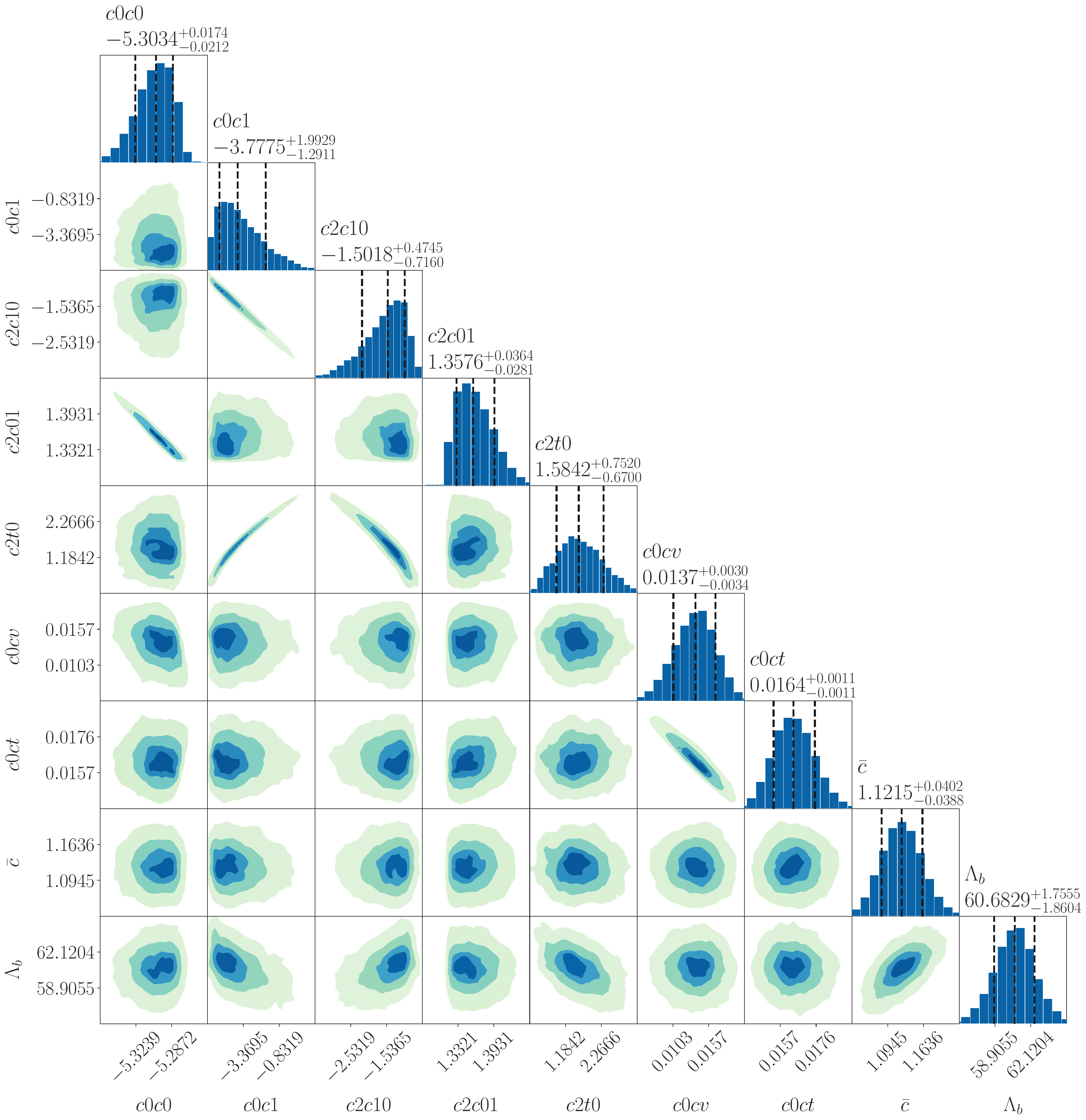}
    \caption{Posterior distributions for $\slashed{\pi}_{\text{WB}}$NLO$_{\text{Red.}}$-1.5 LECs. See Fig.~\ref{fig:2.0.lo.no.trunc.corner} for figure notation.}
    \label{fig:1.5.nlo.s.wave.corner}
\end{figure*}

\begin{figure*}
    \centering
    \includegraphics[width=\linewidth]{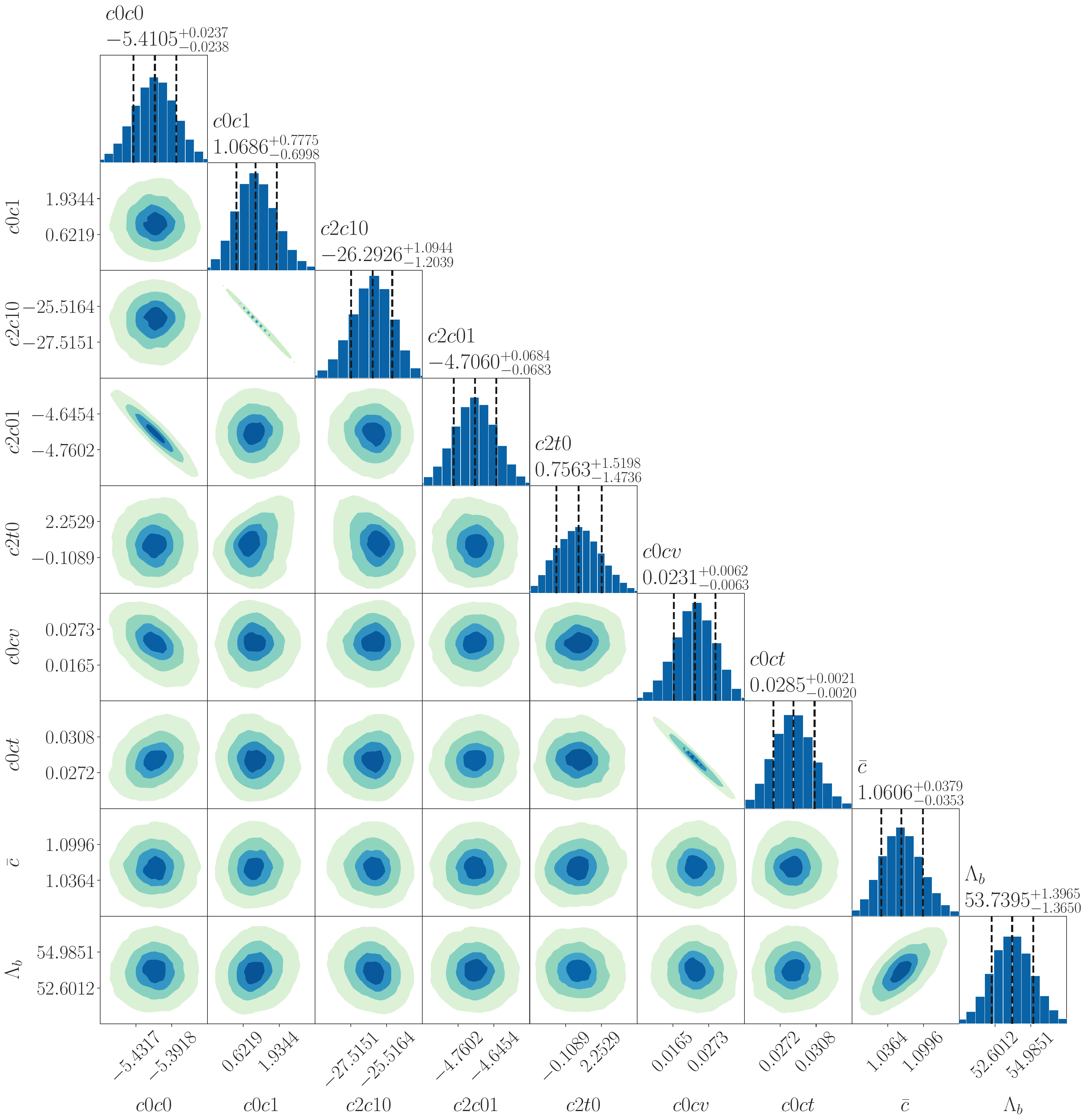}
    \caption{Posterior distributions for $\slashed{\pi}_{\text{WB}}$NLO$_{\text{Red.}}$-2.5 LECs. See Fig.~\ref{fig:2.0.lo.no.trunc.corner} for figure notation.}
    \label{fig:2.5.nlo.s.wave.corner}
\end{figure*}

With an interaction that acts only in even $L$ channels, dominated by $s$-waves, we can perform the same order-by-order calibration that we carried out in Sec.~\ref{subsec:model_calibration}. We label these calibrated interactions as $\slashed{\pi}_{\text{WB}}$(NLO/N3LO)$_{\text{Red.}}$. The posterior distributions for $\slashed{\pi}_{\text{WB}}$NLO$_{\text{Red.}}$-2.0 parameters are shown in Fig.~\ref{fig:2.0.nlo.s.wave.corner}. Figures for the other cutoffs (Figs.~\ref{fig:1.5.nlo.s.wave.corner} and~\ref{fig:2.5.nlo.s.wave.corner}) are given at the end, while figures for the $\slashed{\pi}_{\text{WB}}$N3LO$_{\text{Red.}}$ interactions can be found in the repository~\cite{beft}. In these $p$-wave free distributions, we see similar behavior in the correlation structure for the LECs that we observe for those discussed in Sec.~\ref{subsec:model_calibration} for the full interactions. This behavior is much to be expected, as the channels that the LECs operate in do not change. 

Of more interest are the strong cuts and bimodal structure that exist in the $c0c0$, $c0c1$, $c2c10$, and $c2c01$ LECs as we lower the cutoffs from 2.5 fm to 2.0 fm to 1.5 fm. From our understanding of the data and the construction of the EFT, we can hypothesize possible origins for this behavior. The first point of difference occurs in the change in constraints that the polarized $\Delta \sigma_T$ and $\Delta \sigma_L$ add in calibration. These observables have a dependence on singlet-$s$, $p$-wave, and $s$-$d$ coupled channel physics. However, as we have effectively removed $p$-wave physics from our interaction, and the $s$-$d$ coupled channel is constrained by the deuteron, the polarized $\Delta \sigma_L$ and $\Delta \sigma_T$ now offer constraints on $^1S_0$ channel physics, generating a small secondary mode in the singlet-$s$ LECs $c2c01$. This mode appears almost as a detached tail at the 2.0 fm cutoff and vanishes by the 1.5 fm cutoff. 

If we look at the cuts in the distributions for $c0c0$ and $c2c01$ and the skewness in $c0c1$ and $c2c10$ across the cutoffs, we see that these are cutoff-dependent features. Therefore, we can also attribute some sources for these non-normal distributions from regularization artifacts. As we move from a soft cutoff at 2.5 fm to the hardest cutoff at 1.5 fm, we find a trend of these posterior features becoming more prominent. Since this coincides with a reduction in $d$-wave cutoff artifacts, we can realize that a change in constraints in $d$-waves, and thus in the corresponding LECs, changes with the cutoff. As the cutoff stiffens from 2.5 to 1.5 fm, stronger constraints are placed in the LECs through $d$-waves, as the interaction now must account for more $d$-wave content not provided by artifacts. These change in constraints could contribute to the non-Gaussianity that we observe in $c0c0$, $c0c1$, $c0c10$, and $c0c01$.

\begin{figure}
    \centering
    \includegraphics[width=\linewidth]{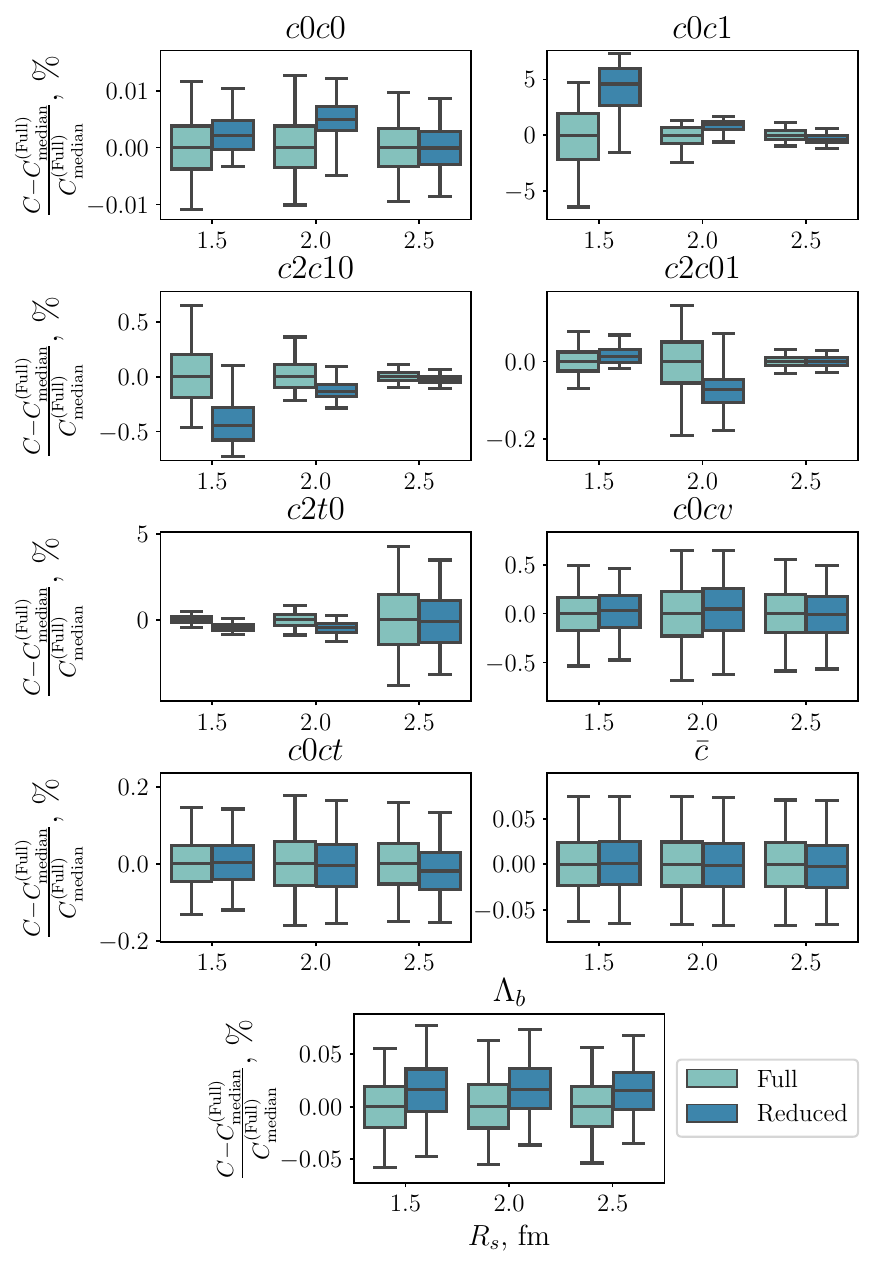}
    \caption{Comparisions of LECs between $\slashed{\pi}_{\text{WB}}$NLO$_{\text{Full}}$ and $\slashed{\pi}_{\text{WB}}$NLO$_{\text{Red.}}$ models where distributions are expressed as percent differences relative to the median value of the $\slashed{\pi}_{\text{WB}}$NLO$_{\text{Full}}$ model for each cutoff. See Fig.~\ref{fig:breakdown_trends} for convention on the box-and-whisker plots.}
    \label{fig:nlo_lec_comparisons}
\end{figure}

However, despite the change in constraints in the $(S,T)=(1,0)$ and $(S,T)=(0,1)$ LECs as we suppress all $p$-wave contributing LECs, we don't see a large shift in the overall values that the remaining LECs take. If we directly compare these LECs between the $\slashed{\pi}_{\text{WB}}$NLO$_{\text{Full}}$ and $\slashed{\pi}_{\text{WB}}$NLO$_{\text{Red.}}$ models, we see significant overlap in their distributions, as shown in Fig. \ref{fig:nlo_lec_comparisons}. Between the full and reduced interactions, we see some difference in the estimated values for the LECs, which occurs due to slight changes in constraints, as discussed in the preceding parts of this section. For $c0c1$, $c2c10$ and $c2t0$, we see a greater than $1\sigma$ difference in the central values between the two types of interactions. However, for all the LECs, we still observe an overlap at the 95\% credibility level, indicating that $p$-waves are not necessary to describe the low-energy data used in the calibration.

Of further note are the estimations of $\cbar$ and $\Lambda_b$ for these models. As we have removed poorly constrained LECs in the construction of this interaction, we do not expect the estimation of these two parameters to change drastically. This is, in fact, what we observe; as most of the data is $s$-wave dominated, the order-by-order corrections remain unchanged regardless of if we remove contributions from LECs that act in the $(S,T)=(0,0)$ and $(S,T)=(1,1)$ channels. This is demonstrated quite clearly in Fig.~\ref{fig:nlo_lec_comparisons}, where the estimation for $\cbar$ and $\Lambda_b$ remain consistent across the cutoffs for the two types of interactions. 

For interactions treated at N3LO, we see similar behavior for $(S,T)=(1,0)$ and $(S,T)=(0,1)$ LECs between the full and reduced interactions. Figures for these interactions at N3LO are in the repository~\cite{beft}.

\subsection{Estimation of Deuteron Properties and Effective Range Parameters}\label{subsec:deut}

Once we have the final posteriors for a particular interaction, we can investigate their predictions of nuclear observables with a full accounting of theoretical uncertainties. In this study, we investigate the deuteron properties and two-body effective range parameters. As discussed in Sec.~\ref{subsec:no_trunc}, we can generate PPDs for the observables. Building on this, we modify Eq.~(\ref{eq:ppd}) with the inclusion of truncation uncertainties,
\begin{align}\label{eq:ppd_full}
    \mathrm{pr}(\mathbf{y}_{\mathrm{th}}|\mathbf{y},\mathbf{x},I) = \int &d\mathbf{a} \,  d\cbar^2 \, d\Lambda_b \, \mathcal{N}\left(\mathbf{y}_{\mathrm{th}},\Sigma_{\mathrm{th}}\right)  \mathrm{pr}(\mathbf{a}|\mathbf{y},I) \times \nonumber \\ 
     &\mathrm{pr}(\cbar^2|\Lambda_b,\mathbf{a},I) \mathrm{pr}(\Lambda_b|\mathbf{a},I)
\end{align}
where we marginalize over $\mathbf{a}$, $\cbar^2$, and $\Lambda_b$. In this procedure, we again rely on the assumption that the truncation errors are normally distributed.

\begin{figure}
    \centering
    \includegraphics[width=\linewidth]{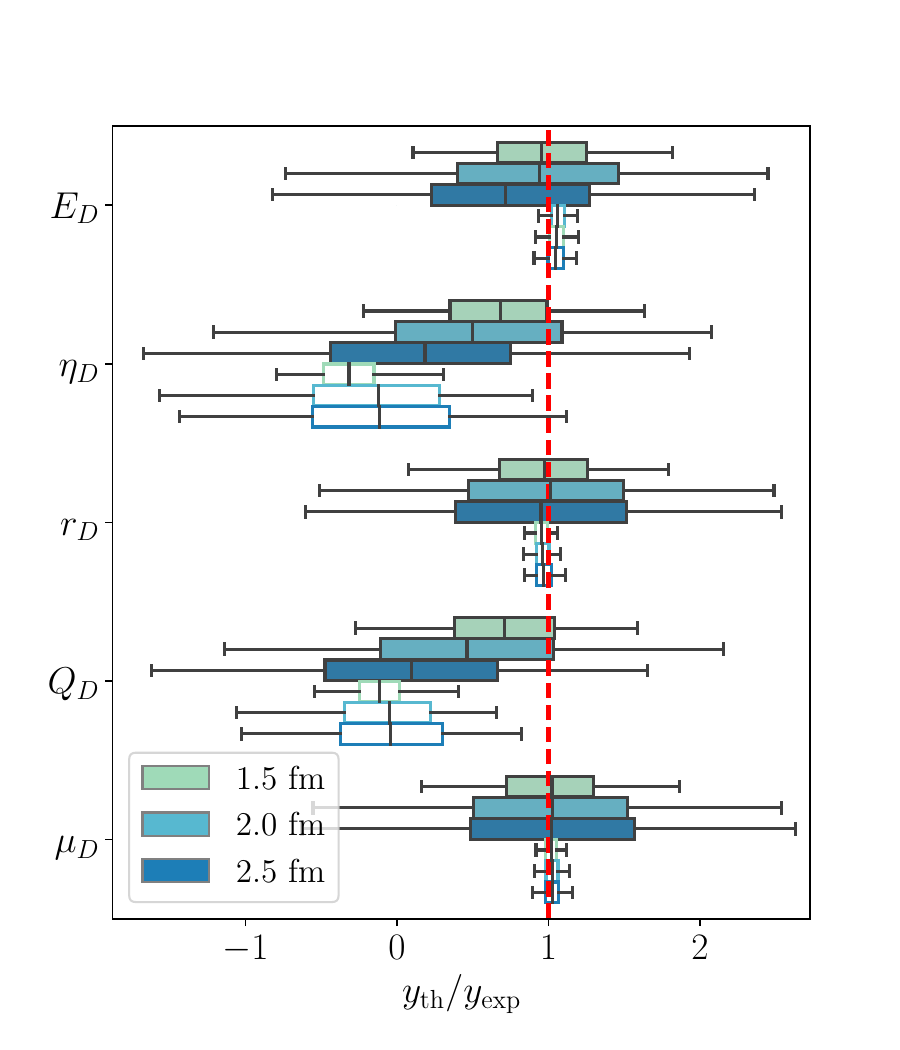}
    \caption{Deuteron properties PPDs for $\slashed{\pi}_{\text{WB}}$(NLO/N3LO)$_{\text{Red.}}$ models normalized to experimental values. NLO interactions are given by the filled boxes, while N3LO interactions are given by empty boxes. See Fig.~\ref{fig:breakdown_trends} for the convention on the box-and-whisker plots.}
    \label{fig:deut_s}
\end{figure}

For deuteron properties, we can generate PPDs for each of the interactions that we sampled in Secs.~\ref{subsec:model_calibration} and \ref{subsec:s_wave_interactions}. 
They are generated by sampling Eq.~(\ref{eq:ppd_full}) for different parameter sets, $\mathbf{a}$. For the covariance that appears in this equation, we model all of the deuteron properties as uncorrelated, as we have no a priori knowledge of how these should be correlated. Further, as $\Sigma_{\mathrm{th}}$ depends on $\cbar$ and $\Lambda_b$, we could choose to sample these parameters or fix these at some value. For simplicity, we fix these values at the median estimated values from the posterior estimation. The PPD for deuteron properties are shown in Fig.~\ref{fig:deut_s} for the $\slashed{\pi}_{\text{WB}}$(NLO/N3LO)$_{\text{Red.}}$ models. An interested reader can find deuteron properties for the $\slashed{\pi}_{\text{WB}}$(NLO/N3LO)$_{\text{Full}}$ models in the repository~\cite{beft}, where similar behavior can be observed.

For each of these interactions, we observe a few general trends. First, for $E_D$, $r_D$, and $\mu_D$, we manage to capture most of the deuteron properties within uncertainty, regardless of whether we include or neglect $p$-waves in the interaction. This is to be expected as these are $s$-wave dominated. Second, for these properties, we can notice a systematic decrease in the uncertainty between NLO and N3LO. At NLO, we see uncertainties of $\approx$150\% on average, which reduces to $\approx$10\% on average at N3LO. These uncertainties are expected from Eq.~(\ref{eq:cor_err}) given the truncation orders and estimated values of $\cbar$ and $\Lambda_b$.  Third, we observe cutoff dependence in these observables, which arises due to regulator artifacts. However, the distributions for different cutoffs agree within uncertainty, and further, the cutoff dependence is smaller at N3LO than it is at NLO. 

Fourth, and in contrast, we can see in Fig.~\ref{fig:deut_s} that  $\eta_D$ and $Q_D$ are systematically lower than the experimental value. This difference is statistically significant at N3LO: while we can describe both of the properties at NLO within uncertainties, we no longer do so at N3LO. This is due to the poor constraints on the $d$-wave content that is present in the scattering data up to $E_{\mathrm{lab}}=5$ MeV. This becomes evident as we look at the ranges these quantities take at N3LO. For the $s$-wave dominant properties ($E_D$, $r_D$, and $\mu_D$), there is a systematic reduction in the uncertainties as we move from NLO to N3LO due to the suppression of the truncation uncertainty between the two orders. However, for $\eta_D$ and $Q_D$, a commensurate reduction does not occur. The truncation uncertainty in Eq.~(\ref{eq:cor_err}) is independent of the scale of the observable when we normalize to experimental data via our choice of $y_{\mathrm{ref}} = y_{\mathrm{exp}}$. Therefore, this qualitative difference in behavior is not due to the relative smallness of these observables, and we deduce that their errors are driven by parametric uncertainty, i.e., LEC uncertainty.

Examining the $\eta_D$ and $Q_D$ results further, we observe that their uncertainty is larger for softer cutoffs. Both these observables are sensitive to $d$-waves, so the larger $d$-wave artifacts of the softer interactions cause a systematic growth in the uncertainty as $R_s$ increases. In contrast, going from NLO to N3LO reduces the regulator artifacts. At N3LO the median value of $\eta_D$ and $Q_D$ becomes very close to zero for all cutoffs: as regulator artifacts diminish, so do $d$-wave components of the interaction. And indeed, the $NN$ data in the calibration's energy range has little $d$-wave content, so it should be the case that our calibrated models have small, and poorly constrained, $d$-wave pieces.

This is an important shift in expectations regarding $\eta_D$ and $Q_D$: calibrated as they are here, our \pionlessEFT\ models should produce estimates for observables that are sensitive to $d$-waves which are consistent with \textit{zero} within uncertainty. And this is exactly what we observe for $\eta_D$ and $Q_D$ in all of our models at both NLO and N3LO. In the case of $Q_D$ the difference between this zero and the experimental value will be made up by a $Q_D$-specific two-body contribution to the electric charge operator that enters at $O(Q^5)$ relative to leading order in Weinberg power counting~\cite{Chen:1999tn,Phillips:2003jz}. This effect is omitted from our impulse approximation calculation of $Q_D$. Furthermore, because uncertainties in Fig.~\ref{fig:deut_s} arise only from the $NN$ interaction, the N3LO truncation uncertainty assigned to $Q_D$ there is $O(Q^6)$ relative to leading, which may explain why the experimental value is not covered by the PPD at N3LO. 

\begin{figure}
    \centering
    \includegraphics[width=\linewidth]{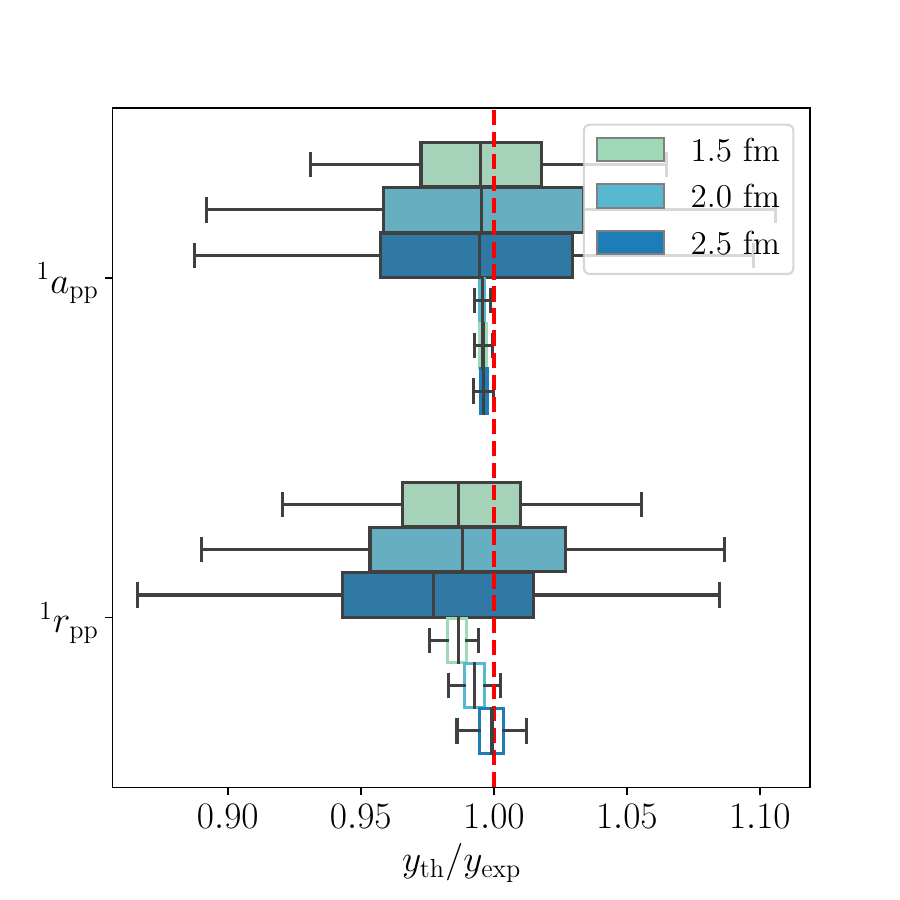}
    \caption{Effective range parameter PPDs for $pp$ scattering for $\slashed{\pi}_{\text{WB}}$(NLO/N3LO)$_{\text{Red.}}$ models normalized to experimental values from Ref.~\cite{Bergervoet:1988zz}. See Fig.~\ref{fig:deut_s} for notation on the interactions and Fig.~\ref{fig:breakdown_trends} for the convention on the box-and-whisker plots. }
    \label{fig:eff_s_pp}
\end{figure}

\begin{figure}
    \centering
    \includegraphics[width=\linewidth]{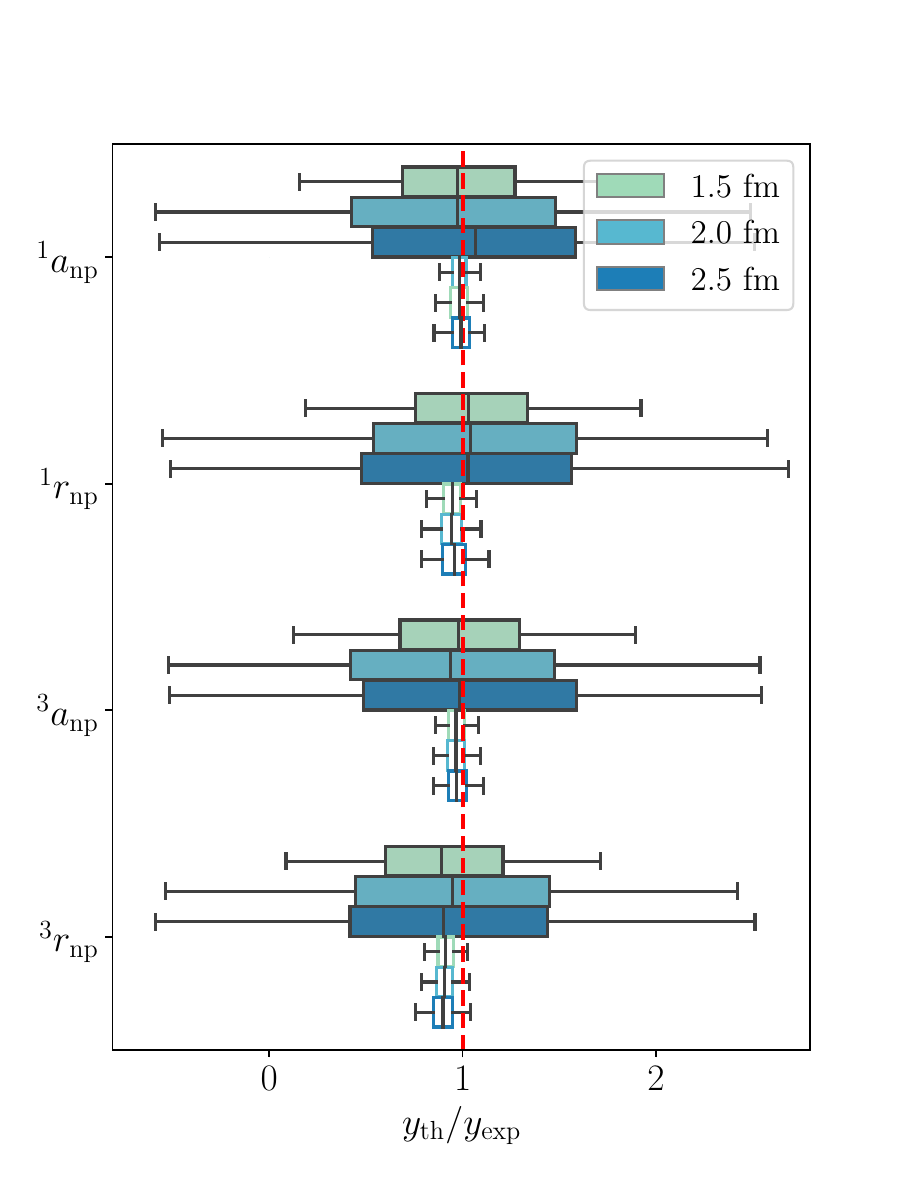}
    \caption{Effective range parameter PPDs for $np$ scattering for $\slashed{\pi}_{\text{WB}}$(NLO/N3LO)$_{\text{Red.}}$ models normalized to experimental values from Refs.~\cite{Dilg:1975zz,Klarsfeld:1984es,Miller:1990iz,Machleidt:2000ge}. See Fig.~\ref{fig:deut_s} for notation on the interactions and Fig.~\ref{fig:breakdown_trends} for the convention on the box-and-whisker plots.}
    \label{fig:eff_s}
\end{figure}

Using the exact same procedure, we generate PPDs for effective range parameters. These are given in Fig.~\ref{fig:eff_s_pp} for the singlet $pp$ effective range parameters and Fig.~\ref{fig:eff_s} for $np$ effective range parameters. In both figures results for the $\slashed{\pi}_{\text{WB}}$(NLO/N3LO)$_{\text{Red.}}$ interactions are shown. These figures show very similar trends to those in Fig.~\ref{fig:deut_s} for $E_D$, $r_D$, and $\mu_D$.
At NLO, we capture the quantities within the large uncertainty band, and that band is then suppressed at N3LO. Figures for the $\slashed{\pi}_{\text{WB}}$(NLO/N3LO)$_{\text{Full}}$ interactions can be found in the repository~\cite{beft}; the full interactions show similar characteristics to the results displayed here for the $s$-wave dominant interactions. As effective range parameters are determined by $s$-wave physics, we capture all of them well, regardless of whether we include or neglect $p$-wave interactions. 

We must also note here the behavior of the uncertainty between the $pp$ effective range parameters and $np$ parameters, as the errors show a marked difference between the two. This is due to the different soft scales used to quantify the truncation uncertainty for the two sets of scattering data. As discussed in Sec.~\ref{subsec:lamb_and_cbar} (see Eq.~(\ref{eq:soft_scales})), the soft momentum scales for the $np$ and $pp$ systems come from different natural scales of the low energy interaction. This creates a difference of roughly a factor of $2$ in the expansion parameter $Q$ once $p < p_{\rm soft}$. This produces vastly different truncation error scales for zero-energy observables like the effective range parameters. $pp$ effective range parameters have a truncation uncertainty that is $\approx 6\%$ (2\%) of the $np$ truncation error at NLO (N3LO). 

For all of these effective range parameters, we compare our estimations to ``experimentally extracted values''. However, it must be remembered that the extractions that we compare are themselves analyses of the scattering data. Thus, while great effort was made to remove the model dependence from the experimental extractions, some may inevitably remain. The effective range parameters that we report in Figs.~\ref{fig:eff_s_pp} and~\ref{fig:eff_s} can be thought of as our own estimations that we have made from the scattering data with model uncertainty included.

\section {Summary}
\label{sec:summary}

We implement the BUQEYE~\cite{BUQEYE} framework to calibrate $\slashed{\pi}$EFT in a Weinberg power counting scheme to $NN$ scattering data. In the analysis, we quantify various sources of uncertainty using our Bayesian framework. First, we demonstrate that we can quantify parametric uncertainty by switching from maximum likelihood fitting to posterior estimation for the LECs. We find that the associated parametric uncertainty produces negligible uncertainties for predicted quantities by constructing the posterior predictive distribution (PPD) using Eq.~(\ref{eq:ppd}). This motivates a full Bayesian model calibration including EFT truncation uncertainties, as prescribed by the BUQEYE framework.

The beginning point for understanding truncation uncertainties is to examine the convergence pattern of the LECs as we include estimates of these uncertainties---and the sensitivity of the LEC extraction to the correlation structure of those uncertainties across lab energies. The results of this exercise are shown in Fig.~\ref{fig:Wesolowski_2_0_fm}. By doing such an exercise, we can observe that the truncation error estimates play a large role in LEC estimation, particularly for LECs that operate in the ${}^3S_1-{}^3D_1$ coupled channel. Further, we find that LECs that operate in $p$-wave channels are under unconstrained at lab energies below 5 MeV since $p$-wave data does not enter the calibration until higher energies than this.

This is an especially important point once we estimate $\Lambda_b$ for our $\slashed{\pi}$EFT implementations. A full EFT calibration to data also involves estimating $\Lambda_b$ and the associated ``number of order one'' $\bar{c}$. In our case it reveals that $\Lambda_b$ may be as small as $50$ MeV, restricting our model calibration to data below $E_{\rm lab}=5$ MeV. LECs that operate in $p$-wave channels are poorly constrained in such a $\slashed{\pi}$EFT calibration, which leads us to examine models that do not contain $p$-wave interactions. We find that in this model the LECs that don't act in $p$-wave channels have PPDs that are consistent with those from the model in which $p$-wave interactions are retained. 

We estimate $\Lambda_b$ for $NN$ potentials at both NLO and N3LO in the EFT and encounter an interesting puzzle: the NLO and N3LO estimations for $\Lambda_b$ disagree, as shown in Fig.~\ref{fig:breakdown_trends}. This disagreement persists even after we use an (admittedly crude) extrapolation of $\Lambda_b$ with regulator scale to remove regulator artifacts (see in Fig.~\ref{fig:breakdown_trends}). While the extrapolation reduces the discrepancy between the NLO and N3LO values of $\Lambda_b$, suggesting part of it is due to 
 Fierz transformation breaking effects, it does not remove it completely. We have proposed solutions for the remaining discrepancy, namely the power counting and the perturbative approximation in Eq.~(\ref{eq:series_expansion}). These mechanisms can be explored in detail in future studies.

Finally, we propagate the parametric and truncation uncertainties that we have now accounted for through PPDs for observables of interest. For the purposes of this study, we choose PPDs for deuteron properties, see Fig.~\ref{fig:deut_s}, and effective range parameters, see Figs.~\ref{fig:eff_s_pp} and~\ref{fig:eff_s}. For these observables, our sets of models produce estimates with large uncertainties at NLO, which are then systematically reduced at N3LO---at least for data that is sensitive to constrained channels. This suppression arises both from the smaller truncation uncertainties at N3LO and from the larger $\Lambda_b$ value obtained at that order. 

In order to fully understand the uses and limitations of PPDs in calculations of many-body observables, future work will apply our $\slashed{\pi}$EFT models to light nuclei. In such systems, we should expect some contribution of $p$-waves to observables of interest. These cases should see a stark difference in the predictive capabilities between the $\slashed{\pi}_{\text{WB}}$(NLO/N3L)$_{\text{Full}}$ and $\slashed{\pi}_{\text{WB}}$(NLO/N3L)$_{\text{Red.}}$ model classes. In particular, the unconstrained $p$-waves in the $\slashed{\pi}_{\text{WB}}$(NLO/N3L)$_{\text{Full}}$ interactions may lead to unphysical PPDs for nuclear binding energies. Work along these lines is in progress.

Looking forward to potentials with $\pi$ and $\Delta$-isobar degrees of freedom, we can readily apply the Bayesian ideas and tools we have used in this work to such interactions. However, the computational expense associated with sampling the posteriors in Eqs.~(\ref{eq:lec_posterior}) and (\ref{eq:total_post}) via MCMC must be accounted for. The main expense in these calculations arises in the solution to the $NN$ Schr{\"o}dinger equation to compute the scattering observables and, hence, the LEC likelihood. In order to investigate EFTs with more degrees of freedom, higher energy $NN$ scattering data must be included in the calibration, and the amount of data grows non-linearly with lab energy. To aid the extraction of high-quality, uncorrelated MCMC samples for the LEC posterior, we can make the calibration more computationally reasonable with the implementation of emulators for the scattering problem. Gaussian process emulators~\cite{Surer:2022lhs, Liyanage:2023nds} and reduced order method emulators~\cite{Furnstahl:2020abp, Melendez:2021lyq, Giuliani:2022yna, Odell:2023cun, Hu:2022} have already demonstrated enormous benefits in speeding up nuclear observable calculations. In particular, emulators have already shown promise when paired with the BUQEYE framework for the calibration of NN LECs in EFTs with pion degrees of freedom~\cite{Svensson:2023twt}.

The primary goal of this study was to apply the
BUQEYE framework to LEC calibration from scattering data, with an eye on ultimately producing high quality $\pi$- and $\Delta$-full interactions that account for truncation uncertainties. To achieve this here we examined the lower-energy $\slashed{\pi}$ regime, where interactions are simpler and the database is smaller. In the process, we uncovered
a discrepancy in $\Lambda_b$ in our $\slashed{\pi}$EFT models between NLO and N3LO. This could be a defect of our choice to implement \pionlessEFT\ by using Weinberg power counting (naive dimensional analysis) for the $NN$ potential. A similar Bayesian analysis should be performed using a different $\slashed{\pi}$EFT power counting---the one developed in Refs.~\cite{Kaplan:1998tg,Kaplan:1998we,vanKolck:1998bw,Birse:1998dk}---to see if that counting produces consistent $\Lambda_b$ extraction across EFT orders. 
Thus implementing the BUQEYE framework for an often-overlooked type of \textit{ab initio} interaction in \pionlessEFT\ suggests a natural and important future study and a possible general way to test the consistency of proposed EFT power countings.

\vspace*{5pt}

\section*{Acknowledgments}
The authors gratefully acknowledge Cole Pruitt for helping with code in this project's infancy, and Alessandro Lovato, Sarah Wesolowski, and Jordan Melendez for insightful discussions. \\
\indent This work is supported by the U.S. Department of Energy (DOE), Office of Science, under the 2021 Early Career Award number DE-SC0022002 (J.B. and M.P.), the FRIB Theory Alliance award DE-SC0013617 (M.P.), DE-SC0021027 (S.P.), DE-FG02-93ER40756 (D.P.), and the NUCLEI SciDAC program under award DE-SC0023495 (J.B., S.P., M.P.) and award DE-FG02-96ER40963 (R.F.). This project is also supported in part by the National Science Foundation (NSF) CSSI program under award No. OAC2004601 (BAND Collaboration \cite{bandframework}) and NSF contract PHY-2209442 (R.F.). J.B., S.P., and M.P. thank the Nuclear Theory for New Physics Topical Collaboration, supported by the U.S. Department of Energy under contract DE-SC0023663, for fostering dynamic collaborations. \\
\indent The calculations were performed on the parallel computers of the Laboratory Computing Resource Center, Argonne National Laboratory. We also extend our thanks to the Washington University in St. Louis Department of Physics for providing computational resources and to Sai Iyer, who made the use of the system possible.

\bibliographystyle{apsrev.bst}
\bibliography{biblio}

\end{document}